\documentclass{article}

\usepackage[final]{neurips_2025}

\usepackage[utf8]{inputenc} 
\usepackage[T1]{fontenc}    
\usepackage{hyperref}       
\usepackage{url}            
\usepackage{booktabs}       
\usepackage{amsfonts}       
\usepackage{nicefrac}       
\usepackage{microtype}      
\usepackage{xcolor}         
 \usepackage{multirow}

\usepackage{amsmath} 
\usepackage{graphicx}
\usepackage{subfig}
\usepackage{mathtools}

\newcommand{\animalA}{\mathsf{A}}
\newcommand{\animalB}{\mathsf{B}}

\DeclareMathOperator{\corr}{\mathsf{Corr}}
\DeclareMathOperator{\rsa}{\mathsf{RSA}}

\DeclareMathOperator{\rdm}{\mathsf{RDM}}

\DeclareMathOperator{\trueAid}{t^{\animalA}}
\DeclareMathOperator{\trueBid}{t^{\animalB}}

\DeclareMathOperator{\sfAid}{s^{\animalA}_{1}}
\DeclareMathOperator{\ssAid}{s^{\animalA}_{2}}

\DeclareMathOperator{\sfBid}{s^{\animalB}_{1}}
\DeclareMathOperator{\ssBid}{s^{\animalB}_{2}}

\definecolor{todoColor}{RGB}{255, 0, 0}   

\definecolor{redColor}{RGB}{200, 0, 0}
\newcommand{\red}[1]{{\color{redColor}#1}}

\definecolor{greenColor}{RGB}{0, 170, 20}
\newcommand{\green}[1]{{\color{greenColor}#1}}

\definecolor{blueColor}{RGB}{20, 40, 200}
\newcommand{\blue}[1]{{\color{blueColor}#1}}

\newcommand{\eadtxt}[1]{{\color{red}\textbf{#1}}}

\definecolor{orangeColor}{RGB}{233, 113, 50}
\newcommand{\orange}[1]{{\color{orangeColor}#1}}

\definecolor{lightorangeColor}{RGB}{251, 227, 214}

\definecolor{skyColor}{RGB}{0, 159, 212}
\newcommand{\skyblue}[1]{{\color{skyColor}#1}}

\definecolor{lightskyColor}{RGB}{202, 238, 251}

\definecolor{pinkColor}{RGB}{160, 0, 160}
\newcommand{\pink}[1]{{\color{pinkColor}#1}}

\definecolor{lightpinkColor}{RGB}{242, 207, 238}

\definecolor{purpleColor}{RGB}{100, 0, 250}
\newcommand{\purple}[1]{{\color{purpleColor}#1}}

\newcommand{\pytorchtnn}{{\href{https://github.com/neuroagents-lab/PyTorchTNN}{\texttt{PyTorchTNN}} }}

\title{Task-Optimized Convolutional Recurrent Networks Align with Tactile Processing in the Rodent Brain}

\RequirePackage{authblk}

\author[*,1]{Trinity Chung}
\author[*,2]{Yuchen Shen}
\author[4]{Nathan C.\;L.\;Kong}
\author[2, 3, 1]{Aran Nayebi}

\affil[1]{Robotics Institute, Carnegie Mellon University; Pittsburgh, PA 15213 }
\affil[2]{Machine Learning Department, Carnegie Mellon University; Pittsburgh, PA 15213 }
\affil[3]{Neuroscience Institute, Carnegie Mellon University; Pittsburgh, PA 15213 }
\affil[4]{Department of Psychology, University of Pennsylvania; Philadelphia, PA 19104}
\affil[*]{
    Equal contribution.

    \vspace{-4pt}
    \texttt{\{trinityc, yuchens3, anayebi\}@cs.cmu.edu};\;\;\texttt{nclkong@sas.upenn.edu}
}

\begin{document}

\maketitle

\begin{abstract}
\label{sec:abstract}
Tactile sensing remains far less understood in neuroscience and less effective in artificial systems compared to more mature modalities such as vision and language.
We bridge these gaps by introducing an Encoder-Attender-Decoder (EAD) framework to systematically explore the space of task-optimized temporal neural networks trained on realistic tactile input sequences from a customized rodent whisker-array simulator. 
We identify convolutional recurrent neural networks (ConvRNNs) as superior encoders to purely feedforward and state-space architectures for tactile categorization. 
Crucially, these ConvRNN-encoder-based EAD models achieve neural representations closely matching rodent somatosensory cortex, saturating the explainable neural variability and revealing a clear linear relationship between supervised categorization performance and neural alignment.
Furthermore, contrastive self-supervised ConvRNN-encoder-based EADs, trained with tactile-specific augmentations, match supervised neural fits, serving as an ethologically-relevant, label-free proxy.

For neuroscience, our findings highlight nonlinear recurrent processing as important for general-purpose tactile representations in somatosensory cortex, providing the first quantitative characterization of the underlying inductive biases in this system. 
For embodied AI, our results emphasize the importance of recurrent EAD architectures to handle realistic tactile inputs, along with tailored self-supervised learning methods for achieving robust tactile perception with the same type of sensors animals use to sense in unstructured environments.

\end{abstract}

\section{Introduction}
\label{sec:intro}
Tactile perception plays an essential role in the manipulation and interpretation of complex environments through active sensing~\citep{lederman2009haptic}.
Animals effectively leverage tactile sensors, such as whiskers, to precisely navigate, forage, and identify objects even in noisy and unstructured conditions~\citep{grant2025whisker}; specifically, contact whiskers deliver critical sensory input for navigation, foraging, and hunting in low-light environments~\citep{ahl1986role}, while water-flow whiskers enable seals to discriminate prey movements from general water currents in similar conditions~\citep{dehnhardt2001hydrodynamic}, and specialized insect hairs respond to wind stimuli, providing information essential for flight stability and agility~\citep{sterbing2011bat}.

Rodents are a common model organism for studying tactile perception in the brain due to the fine experimental control they offer and their ability to discriminate object location, shape, and texture using only their whiskers~\citep{andrew2015low,cheung2019sensorimotor}.
These whiskers, or \emph{vibrissae}, transmit rich mechanical signals to mechanoreceptors at their base, enabling nuanced environmental understanding without direct sensory receptors along their lengths.
In fact, rodent whisking behavior is known to parallel how humans touch objects with their fingertips~\citep{staiger2021neuronal}, such as in the Lateral Motion Exploratory Procedure~\citep{lederman2009haptic}.
Despite significant interest in translating such biological capabilities into robots, artificial systems still struggle to match the tactile perceptual prowess of animals, limiting their functionality in realistic, unstructured environments~\citep{navarro2023visuo}.

There are two primary reasons for this gap: one from the robotics hardware side, and the other from the neuroscience side.
On the hardware side, current bio-inspired whisker sensors for robots face several critical limitations, including substantial hardware complexity when scaling sensor arrays beyond approximately 18-20 whiskers, each requiring individual transducers, significantly increasing wiring and computational demands~\citep{pearson2011biomimetic,assaf2016visual}. 
These sensors also struggle to accurately discriminate multiple simultaneous stimuli such as airflow, direct contact, and inertia, unlike biological whiskers~\citep{simon2023flowdrone}.
Additionally, mechanical discrepancies from biological whiskers--such as reduced sensitivity, limited flexibility, and constrained bending angles--impede precise tactile sensing in dynamic environments~\citep{kent2023identifying,kent2024flow}.

These limitations also apply to anthropomorphic robot hands, which remain in a relatively early stage of development.
For example, besides not being able to sense temperature, good mechanical skin has remained an open challenge in haptics for approximately four decades, primarily due to difficulties in creating artificial skins that maintain realistic deformation during object contact, with current pneumatic approaches proving inadequate to retain shape~\citep{shimonomura2019tactile,dahiya2009tactile}.
While visuotactile camera-based sensors such as GelSight~\citep{yuan2017gelsight} offer a limited haptic solution by providing high-resolution localized surface deformations, they fundamentally differ from human tactile sensing, which actively integrates diverse mechanoreceptor inputs over larger surfaces and multiple contacts~\citep{ward2018tactip}.
Other issues persist with magnetic ~\citep{bhirangi2021reskin}, ~\citep{bhirangi2024anyskin}, which have low spatial resolution and is prone to error from electromagnetic interference ~\citep{bhirangi2024hierarchical}.
These hardware limitations thus make it difficult at the moment to identify robust algorithms for tactile processing that operate on realistic sensory inputs.

On the neuroscience side, despite extensive experimental characterization of rodent somatosensory pathways (e.g.~\citep{armstrong1992flow,kerr2007spatial}), the neural computations underlying precise tactile perception remain poorly understood, primarily due to a scarcity of computational models of this system. 
Rodent whisker sensing involves hierarchical processing, analogous to vision, where sensory signals propagate from primary neurons in the trigeminal ganglion through parallel thalamic pathways, eventually reaching the primary and secondary somatosensory cortices (S1 and S2)~\citep{bosman2011anatomical,moore2015vibrissa}. 
However, relatively few computational approaches have explicitly modeled these neural transformations~\citep{zhuang2017toward}, and none have assessed if any of these models accurately match neural population responses in somatosensory cortex.

To bridge these gaps, we build and systematically evaluate temporal neural networks explicitly trained on biomechanically-realistic force and torque tactile sequences that mice receive, customized from the first complete 3D simulation of the rat whisker array by~\citet{zweifel2021dynamical}. 
We identify convolutional recurrent neural networks (ConvRNNs), particularly IntersectionRNNs, as superior in tactile categorization and neural alignment compared to feedforward (ResNet) and attention-based state-space models (S4 and Mamba). 
Leveraging supervised and contrastive self-supervised learning adapted specifically for tactile data, we demonstrate a direct linear relationship between tactile categorization performance and neural fit, saturating the currently explainable neural variability and surpassing inter-animal consistency benchmarks, with contrastive self-supervised learning serving as an equally neurally predictive, label-free proxy.
Thus, we provide the \emph{first} quantitative characterization of inductive biases required for tactile algorithms to match brain processing.

\section{Related Work}
\label{sec:related}
Task-optimized neural network models have emerged as currently the most quantitatively accurate framework for understanding brain function.
Such goal-driven modeling first successfully explained neural responses in primate visual areas across hierarchical stages~\citep{yamins2014performance,khaligh2014deep}, and subsequently, auditory~\citep{kell2018task,feather2019metamers}, motor~\citep{sussillo2015neural, michaels2020goal}, memory~\citep{nayebi2021explaining}, and language~\citep{schrimpf2021neural} brain areas.
While many of these results are in humans and non-human primates, goal-driven modeling approaches have more recently been extended to mouse visual cortex, demonstrating that shallow architectures trained via contrastive self-supervised learning best capture mouse visual cortical representations~\citep{nayebi2021unsupervised,bakhtiari2021functional}.

Despite these advances, goal-driven computational modeling has not yet been extensively applied to tactile processing, even though tactile sensory systems share hierarchical and recurrent processing features with vision~\citep{felleman1991distributed}.~\citet{zhuang2017toward} provided a foundational goal-driven modeling approach for the rodent whisker-trigeminal system, yet their exploration was limited to relatively simple recurrent architectures and solely supervised categorization loss functions, and they did not compare their models to brain data. 
Here, we significantly broaden the architectural exploration via Encoder-Attender-Decoder parameterization. 
We incorporate sophisticated Convolutional RNN (``ConvRNN'') architectures previously developed by~\citet{nayebi2018task}, which were shown to match primate visual cortex dynamics.
Additionally, we explore advanced temporal models such as State Space Models (SSMs) (e.g. S4~\citep{gu2021efficiently} and Mamba~\citep{mamba}), and Transformers~\citep{haldar2024baku}.
We also employ self-supervised loss functions validated in mouse~\citep{nayebi2021unsupervised,bakhtiari2021functional} and primate visual cortex~\citep{Zhuang2021}, adapted to specifically deal with force and torque inputs.
Finally, we provide the first direct neural comparison to rodent somatosensory cortical responses across all 64 models.

\begin{figure}[b]
  \centering
  \subfloat[]{
  \includegraphics[width=0.7\textwidth]{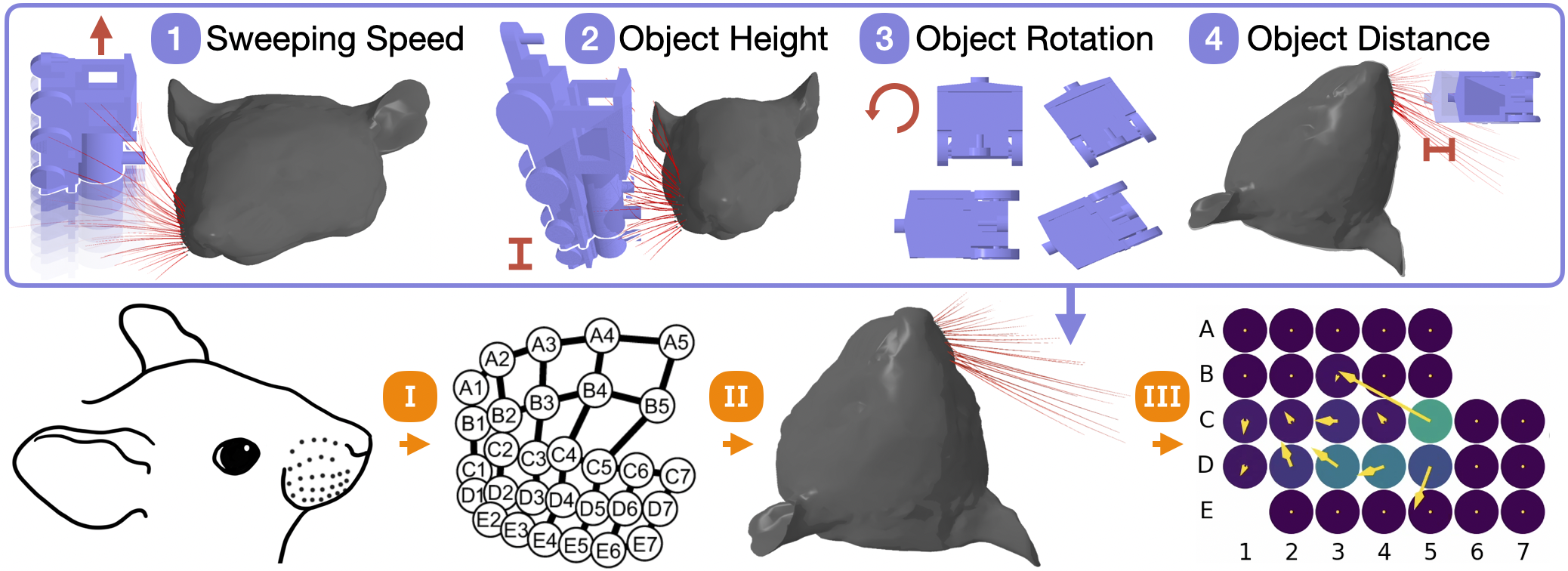}
  \label{fig:dataset_process}
  }
  \subfloat[]{
  \includegraphics[width=0.25\textwidth]{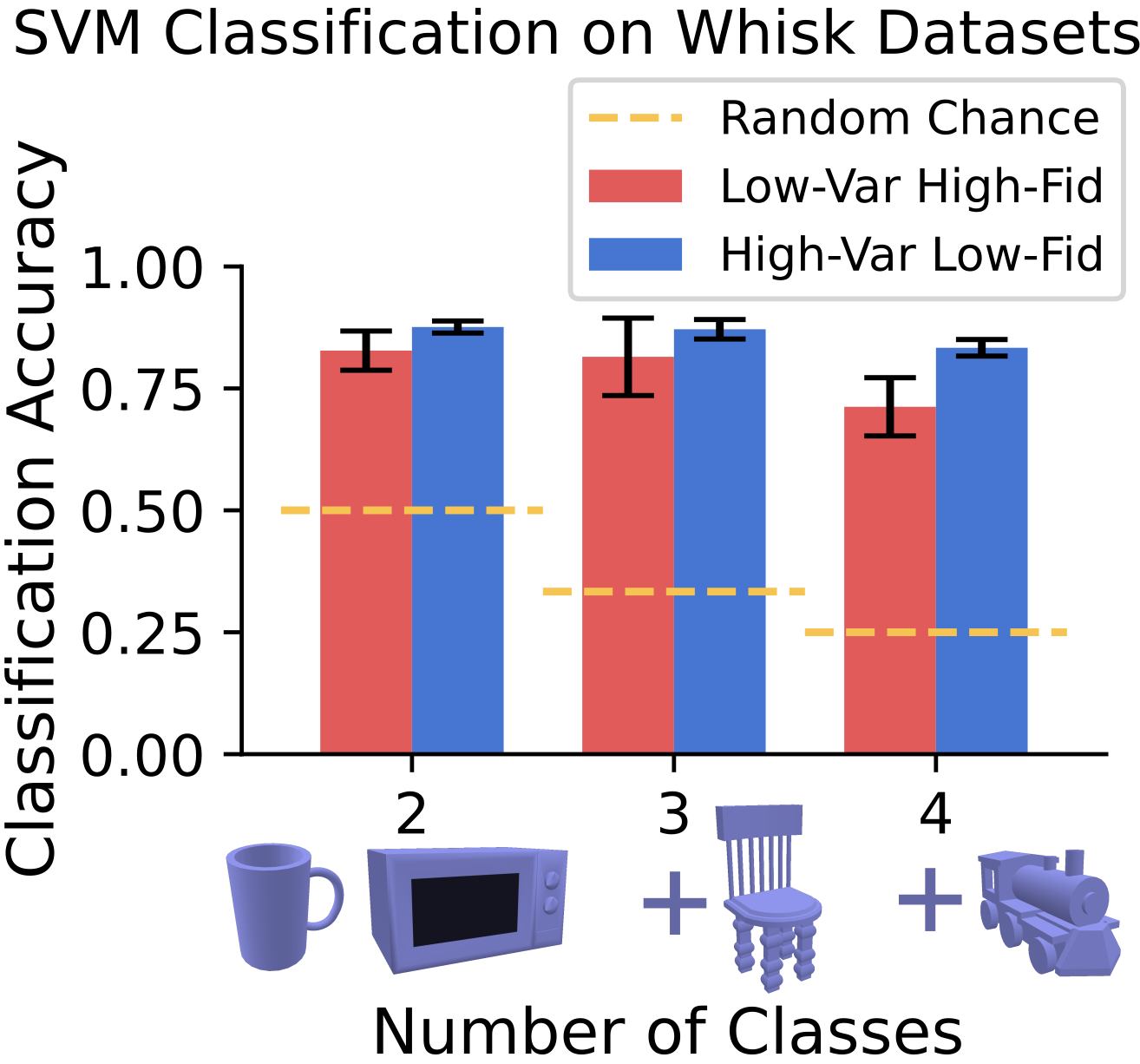}
  \label{fig:svm}
  }
  \caption{
    \textbf{ShapeNet Whisking Dataset.}
      \textbf{(a)}
      \orange{(I)} With average mouse whisker array measurements from \citet{bresee_mice_morphology},
      \orange{(II)} objects are whisked in simulation using WHISKiT  \citep{zweifel2021dynamical}
      resulting in \orange{(III)} force and torque data for sweeping 9981 ShapeNet objects of 117 categories with various sweep augmentations.
      The augmentations vary the \purple{(1)} speed, \purple{(2)} height, \purple{(3)}, rotation, and \purple{(4)} distance of the objects relative to the whisker array.
      We constructed two datasets: a large, low-fidelity set with more sweep augmentations, and a small, high-fidelity set with fewer augmentations
      (see appendix).
      \textbf{(b)}
      An SVM classification on up to 4 different classes of objects (cups, microwaves, chairs, and trains) for the 2 datasets show that the classes are distinguishable (above chance).
  }
  \label{fig:whisk_dataset}
  \vspace{-0.2in}
\end{figure}

\section{Methods}
\label{sec:methods}
All our code for whisker simulation, model training, and neural data analysis is available on GitHub: \url{https://github.com/neuroagents-lab/2025-tactile-whisking}

\textbf{High-variation tactile dataset generation with a biomechanical whisker simulator.}
Given the ongoing development of tactile sensor hardware to match the flexibility of biological tactile sensors such as fingertips and whiskers, we trained our models on synthetic data, focusing on object categorization and self-supervised learning using a high-variation dataset like ShapeNet~\citep{chang2015shapenet}.
This involves the simulated objects interacting with a biomechanically realistic rodent whisker model, WHISKiT Physics~\citep{zweifel2021dynamical}, the first 3D simulation of the rodent's complete whisker system, to provide high-dimensional force and torque inputs. Each of the 30~33 whiskers of an average whisker array~\citep{belli2017variations,belli2018quantifying} is modeled as a chain of rigid conical segments interconnected by torsional springs and actuated according to established equations of motion~\citep{knutsen2008vibrissal}.

Since our objective is to compare models with currently available mouse somatosensory data~\citep{rodgers2022detailed}, we adapted this array to the 30 whiskers of the mouse \citep{bresee_mice_morphology}, arranged as a $5 \times 7$ array with zero padding.
Furthermore, adult mice can exhibit maximal bite forces of approximately 8-10 Newtons~\citep[Table 1]{freeman2008simple}, and since facial tolerance to applied forces would realistically be a fraction of this maximal bite force, our chosen simulation clipping range of $\pm$1000 milli-Newtons ($\pm$1 N) remains comfortably within biologically plausible limits.

We created a whisking dataset for the shape categorization task, generated by passively brushing objects against the whisker array.
Our primary consideration here is to generate a high-variation dataset by which to strongly constrain the representations that are learned, allowing us to more effectively simulate evolutionary pressures, in line with the ``Contravariance Principle''~\citep{cao2024explanatory2} of goal-driven modeling.
On each of the 9981 different objects from ShapeNet, we apply several combinations of sweep augmentations, including adjusting the object sweeping speed, height, rotation, and distance relative to the whisker array (Fig. \ref{fig:whisk_dataset}).
These augmentations on our passive whisker sweeps can be considered to produce data that is isomorphic to performing active whisking in a systematic way.
The 117 object categories are selected based on the work of \citet{zhuang2017toward} to ensure reasonably distinguishable classes.
We constructed two datasets: one large dataset replicating the \citet{zhuang2017toward} data with 288 randomized sweep augmentations and a physics simulation step rate of 110 Hz (High-Variation/Low-Fidelity), and another with 16 sweep augmentations and a higher simulation step rate of 1000 Hz (Low-Variation/High-Fidelity).
See the Appendix~\ref{sec:Dataset} for detailed sweep augmentation procedures.
Although higher temporal resolution is available from the simulation, we extract 22 timesteps for both datasets, corresponding to the average whisking frequency of 20 Hz in rodents~\citep{sofroniew2015whisking}.

\textbf{Encoder-Attender-Decoder (EAD) temporal model search parameterization.} Tactile recognition is performed by the somatosensory cortex, which, though less understood than vision, exhibits hierarchical processing and long-range feedback connections~\citep{lederman2009haptic,navarro2023visuo}. 
These basic anatomical insights motivate our exploration of hierarchical, temporal neural network (TNN) models such as ConvRNNs shown previously to match primate visual cortex dynamics~\citep{nayebi2018task,nayebi2022}, SSMs~\citep{gu2021efficiently}, Transformers~\citep{haldar2024baku}, and ResNets~\citep{He2016} as a feedforward control.
We provide our library \pytorchtnn which enables large-scale exploration of integrating multiple TNN modules that are also combined by long-range feedback with feedforward connections.

\begin{figure}
  \centering
  \subfloat[]{
      \includegraphics[height=5.4cm]{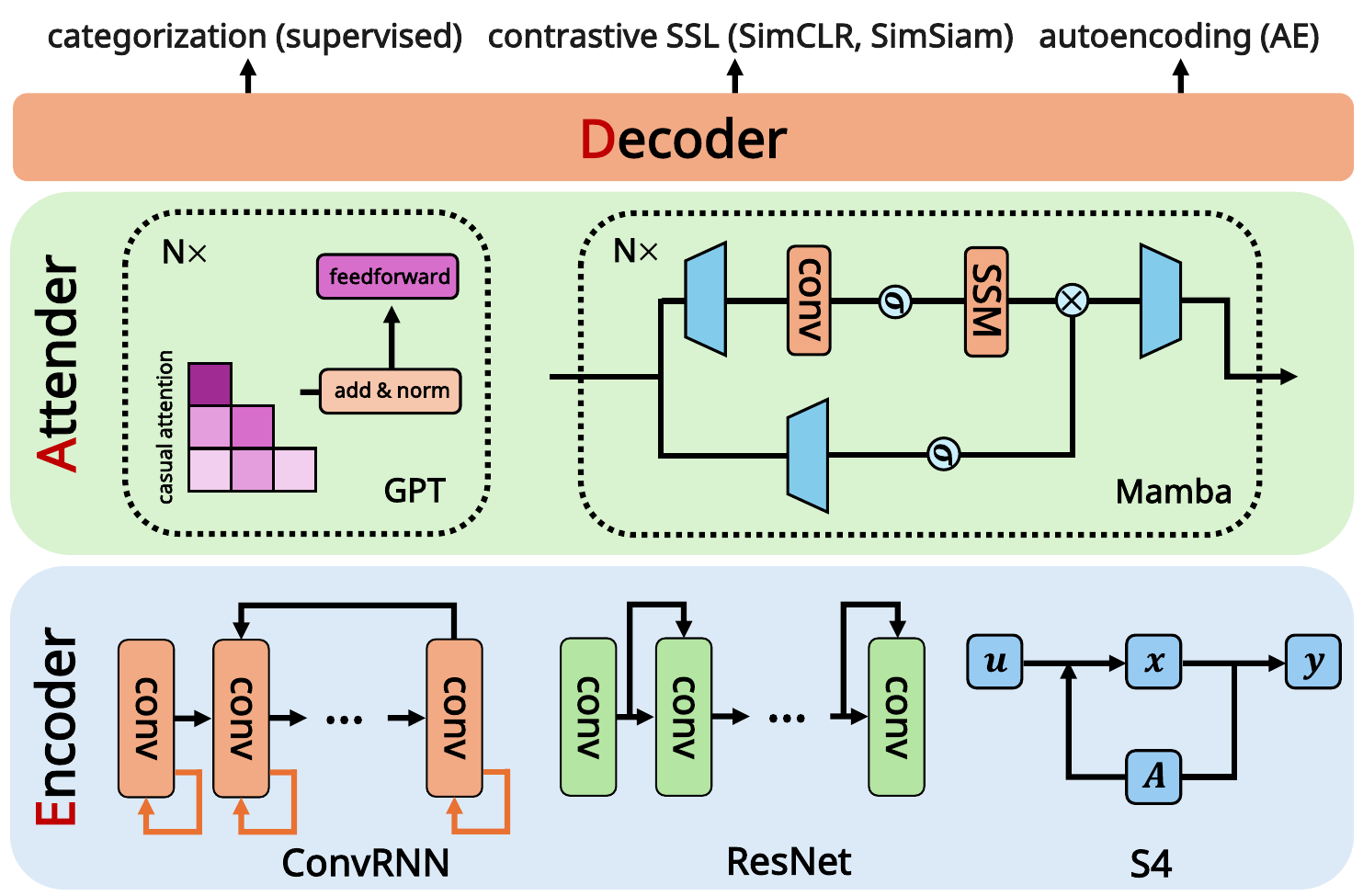}
    \label{fig:ead_architecture}
  }
  \subfloat[]{
      \includegraphics[height=5.4cm]{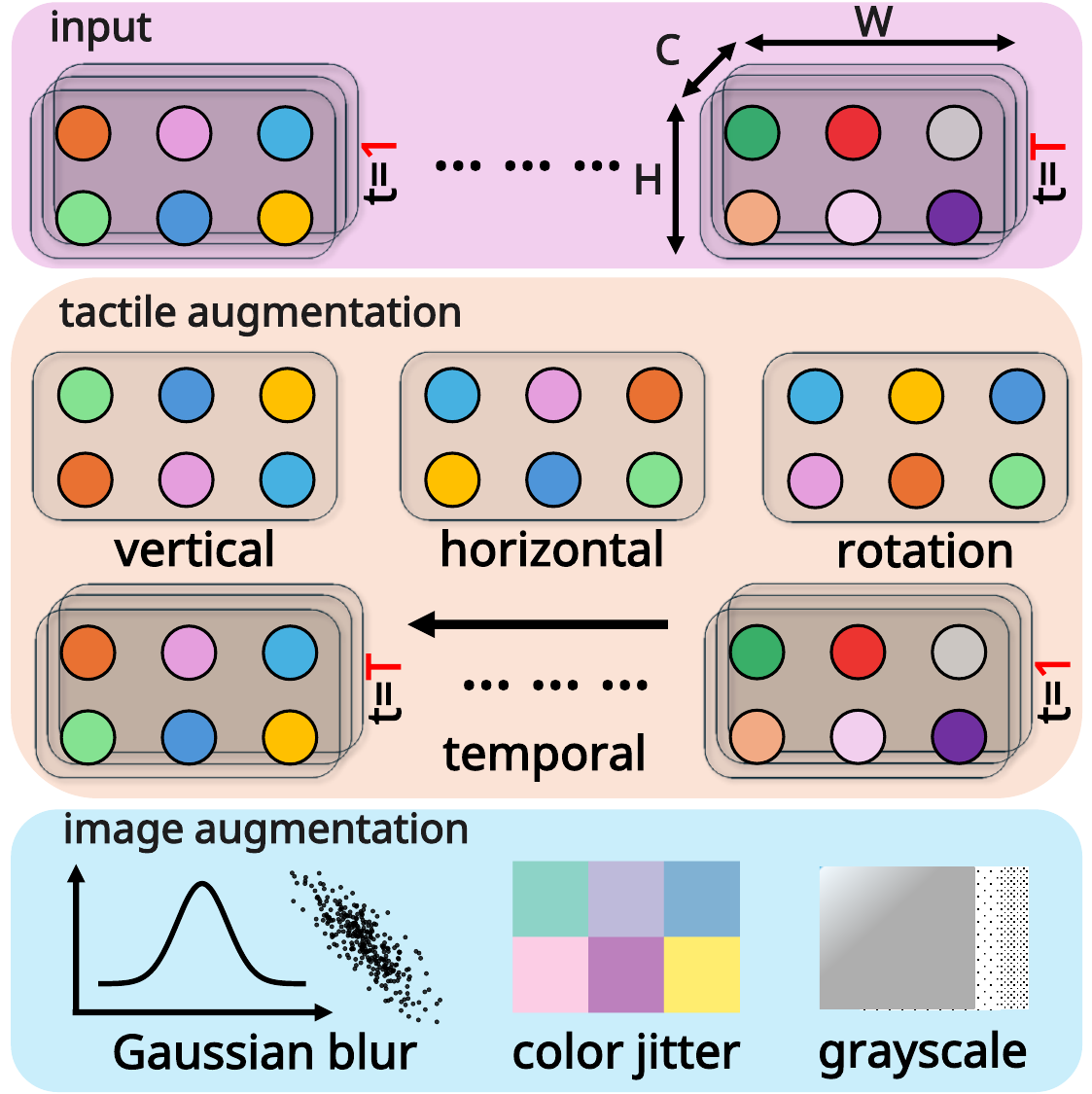}
    \label{fig:ead_transforms}
  }
  \caption{
  \textbf{(a)}
  \eadtxt{E}ncoder-\eadtxt{A}ttender-\eadtxt{D}ecoder (\eadtxt{EAD}) architecture, with task objectives being supervised categorization, self-supervised learning (SimCLR, SimSiam, autoencoding).
  The ConvRNN encoder includes \orange{\textbf{self-recurrence}} at each layer where we vary different RNNs.
  \\
 \textbf{(b)} Types of data augmentations applied to SSL models.
  Given a temporal \pink{\textbf{tactile input}} over time $T$, our \orange{\textbf{tactile augmentation}} vertically, horizontally, temporally flips, and rotates the features, while traditional \skyblue{\textbf{image augmentation}} introduces Gaussian noise, color jitter, and grayscale.}
  \label{fig:EAD_ssl_transform}
  \vspace{-0.2in}
\end{figure}

These considerations produce a rather large search space of model architectures.
Therefore, to be able to effectively search the space of TNN models systematically, we came up with an Encoder-Attender-Decoder (EAD) parameterization, schematized in Fig.~\ref{fig:ead_architecture}.
Convolutional recurrent and state-space layers (such as ConvRNNs and S4) are particularly effective for encoding smooth and compressible temporal signals, given their recurrent smoothing properties and linear-time complexity. 
In contrast, transformers and other attention-like architectures (e.g., Mamba) excel at processing temporally sparse or irregularly informative inputs, as they independently weight each timestep.
Given that tactile inputs from force and torque sensors provide temporally smooth signals, encoder \textbf{(E)} layers closer to these inputs naturally benefit from convolutional and recurrent mechanisms that integrate information locally over time, effectively reducing redundancy and noise. 
In contrast, higher-level temporal aggregation \textbf{(A)}, which must selectively integrate meaningful signals across longer and potentially irregular timescales, is better served by attention-based architectures like Transformers or Mamba, which can dynamically weigh distinct timesteps based on their informational content.
Finally, the decoder \textbf{(D)} is either the classification layer for supervised learning, or similarly producing self-supervised features for contrastive learning or autoencoding.
The EAD paradigm flexibly combines these complementary modules, enabling an effective and structured exploration of the temporal model space for tactile processing.

Specifically, we explore the following EAD combinations (visualized in Fig.~\ref{fig:ead_architecture}): 1) Encoder: Zhuang's recurrent model~\citep{zhuang2017toward}, ResNet~\citep{He2016}, S4~\citep{gu2021efficiently}; 2) Attender: Transformer~\citep{haldar2024baku} (e.g., GPT), Mamba~\citep{mamba}, None (i.e., attender-free); 3) Decoder: MLP. We further investigate different variants of Zhuang's model by replacing the recurrent layers in the ConvRNNs with UGRNN~\citep{Collins2017}, IntersectionRNN~\citep{Collins2017}, LSTM~\citep{Hochreiter1997}, and GRU~\citep{cho2014learningphraserepresentationsusing}, and we present the update rules for the variants in Appendix~\ref{sec:Model}. 
In the \pytorchtnn library, each time step in ConvRNNs corresponds to a single feedforward layer processing its input and passing it to the next layer, in contrast to treating each entire feedforward pass from input to output as a single integral time step, as is normally done with RNNs~\citep{Spoerer2017}. This implementation of temporal unrolling parallels biological systems, where stimuli are sequentially processed from one cortical layer to the next.

\textbf{Model Training and Tactile Augmentations.} Besides the supervised categorization objective, we also consider self-supervised learning (SSL) losses: SimCLR~\citep{chen2020simple}, which learns robust representations via distinguishing the embeddings of augmentations of one image from other images, SimSiam~\citep{Chen2020Siam}, which maximizes the similarity between the embeddings of two augmentations of the same image, and autoencoding (AE)~\citep{olshausen1996emergence}, where we use a 3-layer deconvolution network to reconstruct an image from its sparse latent representation.

We use a train/validation/test split of 80/10/10\%, and report the top-5 test accuracy. 
For supervised learning, we train models for 100 epochs and test on the checkpoint saved with the highest validation accuracy. 
For SSL objectives, we train for 100 epochs and save the model with the lowest validation loss, then continue training for another 100 epochs with the checkpoint frozen and a learnable linear layer on top of it.
To stabilize training, we also consider adding layer norm~\citep{ba2016layernormalization} to the ConvRNNs as an alternative when training was unstable. 
All experiments are conducted on NVIDIA A6000 GPUs, and one set of model search experiment takes 8$\sim$16 hours.

In addition to the various \textit{sweep augmentations} used to generate the initial dataset, we apply tactile augmentations at training time as a cheap way to generate more training data.
We illustrate the tactile input, our tactile augmentations, and traditional image augmentations for SSL in Fig.~\ref{fig:ead_transforms}. 
Given a temporal tactile input over time $T$, the proposed augmentation either vertically flips, horizontally flips, rotates, or reverses the input over time. 
Although the variable whisker lengths mean that these operations are not completely physically accurate, our tactile augmentations still meaningfully mimic flipping/rotating the whisker array or reversing the direction of motion.
We considered temporal masking as an additional augmentation, but found that this did not significantly improve task performance (Appendix~\ref{sec:Additional}, Tab.~\ref{tab:masking}).
Applying traditional image augmentations like color jitter or gray scale cause the models to fail to train, which provides evidence that the choice of augmentations should represent operations relevant to the given modality.


\textbf{Neural Data Comparison.} 
We use the neural dataset from~\citet{rodgers2022detailed}, which recorded neural population activity across superficial (L2/3) to deep (L6) layers of the barrel cortex 
as the mice used their whiskers to perform a simple 2-object (convex/concave) classification task.
Note that this low number of distinct stimuli is a critical limitation which we address in the Discussion section.
After filtering for valid trials, the neural data we used contains a total of 999 neural units across 11 mice.
The time window of each trial is clipped to the time of first whisker contact with the object until the time the mouse makes its decision (about 1-2 seconds).
The neural response is binned into intervals of approximately 45–50 ms, corresponding neatly with the 15–20 Hz whisking frequency typical of rodents (50 ms per cycle at 20 Hz~\citep{sofroniew2015whisking}), and aligning specifically with our model's integration scheme of five sub-timesteps (\textasciitilde 9 ms each) per timestep.
This choice provides a temporal resolution optimal for capturing detailed neural activity during whisker-object contact periods.

We evaluate neural alignment using Representational Similarity Analysis (RSA), a parameter-free approach that compares the pairwise dissimilarity structure of model and neural population responses across the same set of stimuli.
RSA compares the pairwise dissimilarity structure of neural responses and model activations to a common set of stimuli, allowing direct comparison of representational geometry without requiring additional parameter fitting~\citep{kriegeskorte2008representational}.

To ensure only reliable neurons are included, we compute split-half internal consistency of the neural responses across trials and retain only neurons with a Spearman-Brown–corrected reliability greater than 0.5.
All RSA evaluations are thus conducted on this filtered subset of self-consistent neurons, and corrected by this internal consistency.
Further details about this procedure are included in Appendix~\ref{sec:Neural}.

The noise-corrected RSA Pearson's $r$ score is also computed between one mouse to every other mouse, and then averaged this score across all mice to obtain the mean animal score which will serve as the baseline for model-brain evaluation, to account for inherent animal-to-animal variability (which we denote as ``a2a'' in the barplots).
Thus, we want our models to match the brain, at least as well as animals do to each other.

Next, we replicated the experimental setup of the barrel cortex dataset ~\citet{rodgers2022detailed} \emph{in silico}.
Fig.~\ref{fig:neural_fits_stimuli} shows the 6 different stimuli used in the experiment, which are concave/convex objects at three different distances (near, medium, far) from the whisker array.
With 3D model reproduction of the concave/convex objects in the experiment, active whisking is simulated under the 6 stimuli using a real recorded whisking trajectory \citep{bresee_mice_morphology} to generate the model input.

We then obtain the neural alignment score for models by computing the maximum, across all layers, of the median RSA score between each layer's representation and the neural responses, averaged across mice.
The standard error is calculated across the RSA Pearson's $r$ value from the model to each mouse.

\begin{figure}[b]
  \centering
  \includegraphics[width=\textwidth]{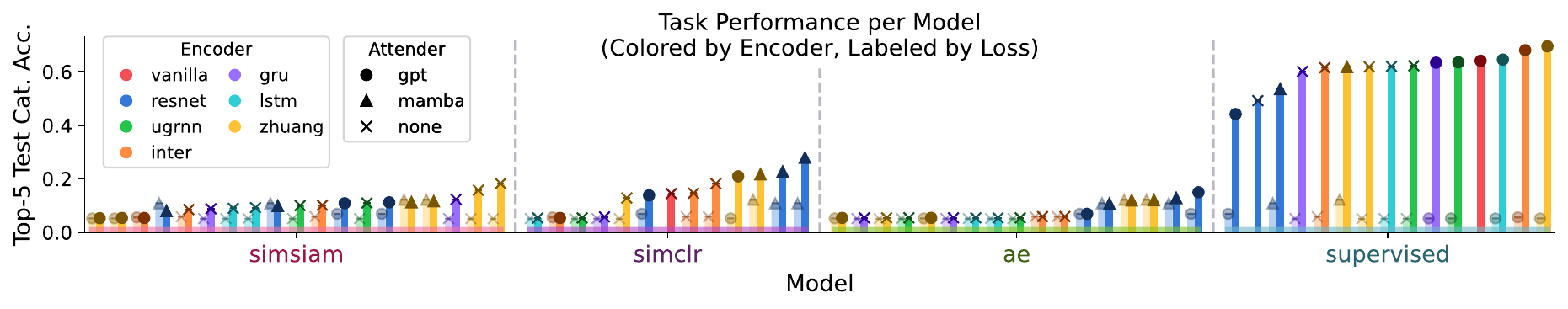}
  
  \caption{\textbf{Tactile Categorization Accuracy.} 
  The lighter-colored left bar represents the randomly initialized version for every model.
  The best-performing model is Zhuang+GPT+Supervised (rightmost yellow bar). Models with the encoder being S4 are excluded as the training losses explode before the first epoch is finished.
  } 
  
  \label{fig:task_performance}
  \vspace{-0.1in}
\end{figure}

\section{Results}
\label{sec:results}

\textbf{Validation of Sensor and Dataset via SVM Decoding.}
We first verified the basic discriminability of the whisker array dataset using support vector machine (SVM) classifiers
on the High-Variation/Low-Fidelity and Low-Variation/High-Fidelity datasets (Fig.~\ref{fig:svm}).
Both showed strong decoding performance, exceeding chance and remaining robust even with increased task difficulty from more object categories. 
The comparable results across sampling rates confirm that the smaller Low-Var./High-Fid. dataset retains sufficient discriminative tactile information, allowing for efficient model training without compromising performance.

\textbf{ConvRNN Encoders Outperform Feedforward and Attention-based Architectures for Tactile Categorization.}
We next systematically evaluated the task performance of the EAD architectures trained on tactile force/torque sequences.
Overall, we found that across EAD architectures, the choice of encoder (E) was quite important for tactile recognition.
Specifically, ConvRNN encoders, especially the IntersectionRNN~\citep{Collins2017}, surpassed the purely feedforward ResNet18 and SSM (S4) encoders, in supervised tactile categorization tasks (Fig.~\ref{fig:task_performance}).
Additionally, models trained with our custom force-and-torque-specific contrastive self-supervised learning (SSL) augmentations (Fig.~\ref{fig:EAD_ssl_transform}b) outperformed untrained networks of the same architecture, and those trained with standard image-based augmentations that involve Gaussian blur and color jittering~\citep{chen2020simple} did not train with the best architecture despite hyperparameter tuning, demonstrating the importance of tailored tactile augmentations for enhancing task performance (Fig.~\ref{fig:task_performance}).


\begin{figure}
\vspace{-0.2in}
  \centering
  \subfloat[
  ]{
    \includegraphics[width=0.7\textwidth]{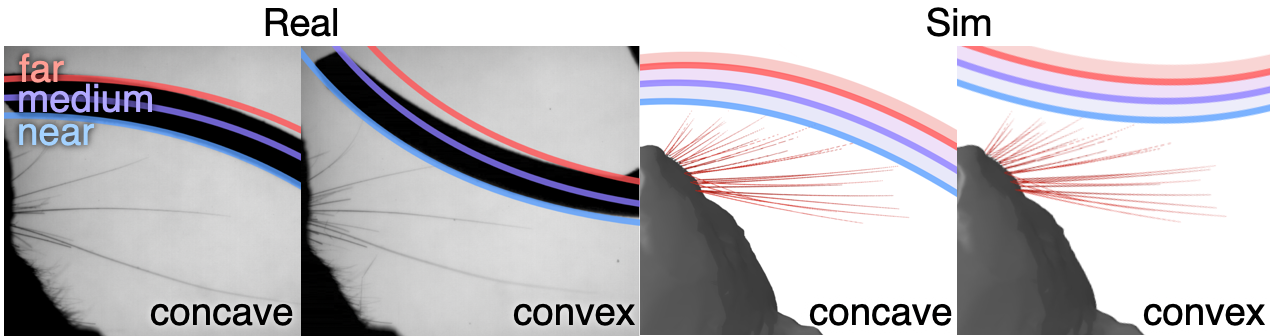}
  \label{fig:neural_fits_stimuli}
  }
  \\
  \subfloat[]{
      \includegraphics[width=\textwidth]{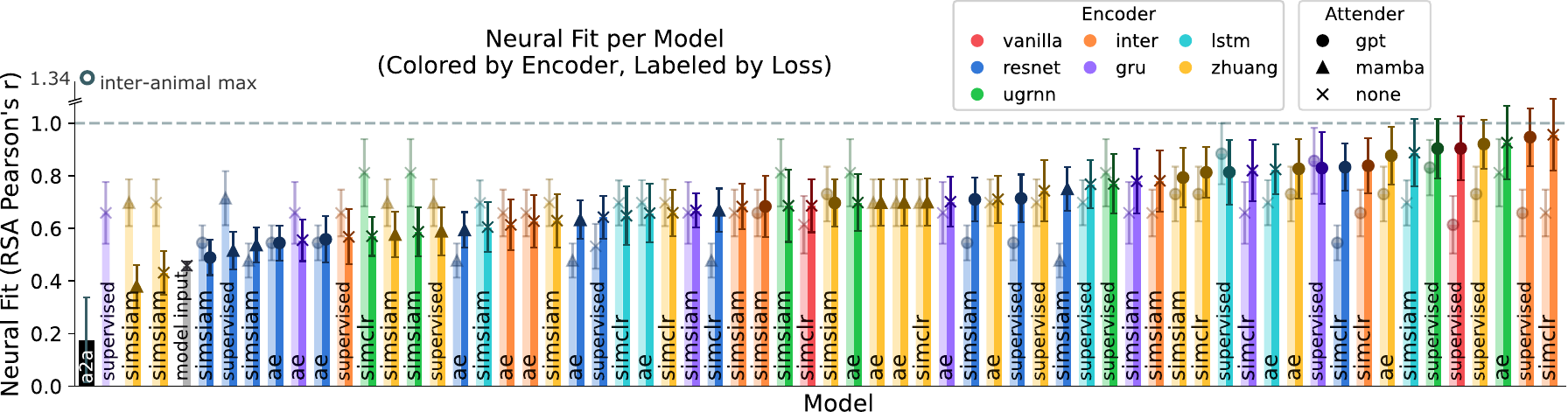}
  \label{fig:neural_fits_doublebar}
  }
  \caption{
    \textbf{Model Neural Evaluation.}
      \textbf{(a)}
          We use six different stimuli  
      (concave/convex $\times$ \textcolor{blue}{near}/\textcolor{purple}{medium}/\textcolor{red}{far})
      replicating the conditions in the mouse neural dataset in simulation.
      (Real images were taken from video recordings in neural dataset \citep{rodgers2022detailed}).
      \textbf{(b)}
      Comparison of neural fit (noise-corrected RSA Pearson's $r$) across models. The mean animal-to-animal score is 0.18 and the maximum between all pairs of animals is 1.34. The leftmost ``a2a'' bar represents the \textit{mean} animal-to-animal neural consistency score.
      The lighter-colored left bar represents the randomly initialized version for every model.
  }
  \label{fig:neural_fits}
  \vspace{-0.2in}
\end{figure}

\textbf{Importance of Architectural Inductive Bias.}
We note that the raw model input (the tactile force/torque sequences obtained from actively whisking on the 6 stimuli objects) achieves a correlation of 0.46 (Fig.~\ref{fig:neural_fits_doublebar}), and untrained randomly initialized models also achieve moderately positive correlation to the mouse neural data.
Randomly initialized networks are highly non-random functions as they are architectures selected to perform the task well when trained. 
Therefore, architecture matters a lot, and this is a well-noted phenomenon in NeuroAI across brain regions~\citep{yamins2014performance, nayebi2021unsupervised, schrimpf2020integrative}.
By doing this comparison, it allows us to isolate the contributions of the architecture vs. loss function for every pair of (architecture, loss function) tuples.
The models that saturate our noise ceiling (bars on far right in Fig.~\ref{fig:neural_fits_doublebar}) are noticeably improved when trained. 

\begin{figure}
  \centering
\subfloat[]{
  \includegraphics[height=4cm]{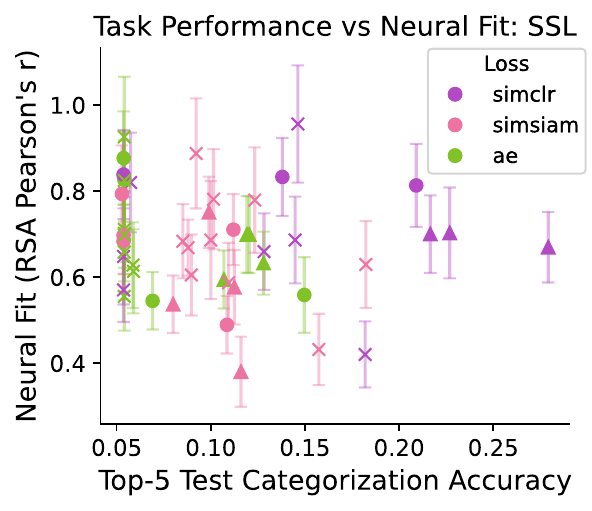}
  \label{fig:neural_eval_ssl}
}
\subfloat[]{
  \includegraphics[height=4cm]{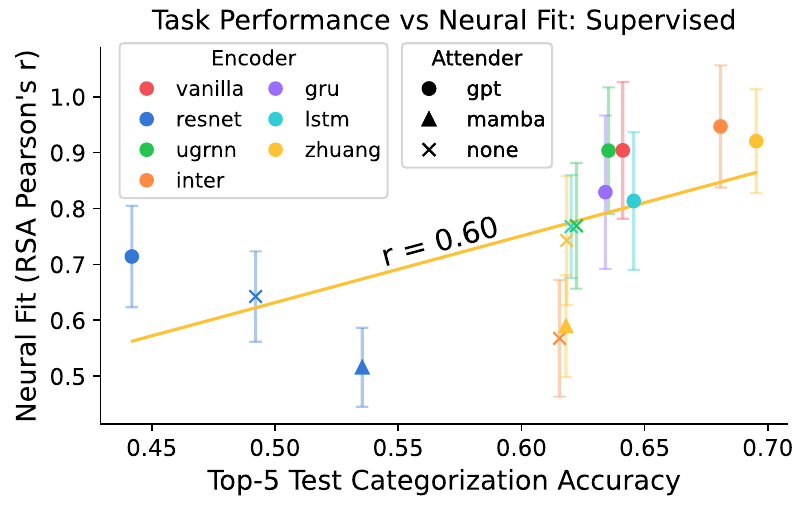}
  \label{fig:neural_eval_supervised}
}
\subfloat[]{
  \includegraphics[height=4cm]{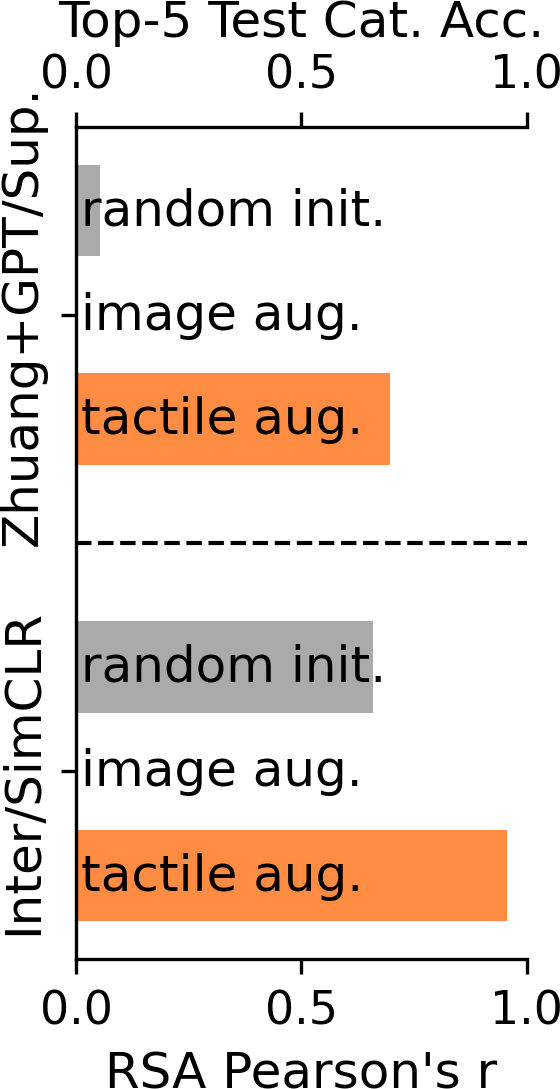}
  \label{fig:neural_eval_augmentations}
}
  \caption{
  \textbf{Comparing Task Performance and Neural Fit.}
  \textbf{(a)}
  The task performance of SSL models are about one order of magnitude below the performance of supervised models, yet are able to achieve comparable neural fit.
  \textbf{(b)}
  For supervised models, we observe a trend of better task performance leading to increased neural correspondence.
  Plotting a best fit line, we find the correlation $r=0.59$.
  \textbf{(c)}
  The tactile augmentations were effective in improving both the neural fit and task performance. The models were unable to be trained with image augmentations. 
  }
  \label{fig:neural_eval}
  \vspace{-0.1in}
\end{figure}

\textbf{ConvRNN Encoders Saturate Explainable Neural Variance in Rodent Somatosensory Cortex.}
To assess biological realism, we compared internal model representations to neural recordings from rodent whisker somatosensory cortex. 
All trained EAD models outperformed raw sensor inputs in neural alignment, underscoring the importance of nonlinear temporal processing in modeling brain-like tactile representations
(Fig.~\ref{fig:neural_fits_doublebar}).
In fact, the best models saturated \emph{all} of the held-out explainable neural variance--\emph{without} fitting any additional parameters--even when tested on entirely novel objects under substantially different experimental conditions (active whisker sensing rather than passive contact, as used in training). 
Remarkably, the neural predictivity exceeded the average inter-animal neural consistency (leftmost black ``a2a'' bar in Fig.~\ref{fig:neural_fits_doublebar}), thereby robustly passing the NeuroAI Turing Test on this dataset~\citep{feather2025brain}.
Among these and consistent with the categorization results, EAD models with ConvRNN encoders consistently provided better neural fits compared to feedforward (ResNet) and SSM-based (S4) encoder architectures, underscoring the biological plausibility of recurrence in modeling tactile processing.
In fact, we saw a strong linear trend between tactile supervised categorization performance and neural fit ($r=0.59$, Fig.~\ref{fig:neural_eval_supervised} ), with the model layers that best predict the tactile neural responses being closest to the decoder layer (Fig.~\ref{fig:layer_neural_scores}).
Although the number of task-optimized parameters matters for both categorization test set performance and neural fit generalization (Fig.~\ref{fig:model_params}), they are not the whole story, as the SimCLR-trained EAD with the IntersectionRNN encoder best matches the neural data with far fewer parameters ($\sim 3.80\times 10^7$ parameters) than its supervised counterpart with a GPT-based attender ($\sim 6.38\times 10^7$ parameters).

\textbf{GPT-based Attenders Provide Modest Improvements in Task Performance and Neural Alignment.}
In addition to varying the encoder (E) layer of the EADs, we investigated different temporal aggregation (``Attender'') schemas. 
We observed that GPT-based Attenders modestly outperformed Mamba-based and no-attention controls in both supervised task categorization and neural alignment (Fig.~\ref{fig:neural_eval_supervised}). 
Although these improvements were subtle, the consistency of the result suggests that incorporating some form of attention downstream of the ConvRNN encoder is beneficial, particularly for our highest-performing neural models.
This result suggests the prediction that attention-like mechanisms may be implemented in somatosensory cortex through selective modulation of hierarchical processing pathways, such as those from primary sensory neurons in the trigeminal ganglion through the thalamus and subsequently into primary and secondary somatosensory cortical areas (S1 and S2), which could be validated by experiments involving targeted perturbations or optogenetic manipulation of these specific pathways during tactile discrimination tasks.

\textbf{Self-Supervised SimCLR Training Matches Supervised Models and Serves as an Ethologically-Relevant Label-Free Proxy.}
We compared neural alignment achieved by supervised training against SSL methods, adapted with tactile-specific force-and-torque augmentations. 
This comparison is necessary because discriminating among 117 human-recognizable shape classes is not directly ethologically relevant for rodents. 
Nevertheless, ShapeNet's extensive diversity of 3D objects provides strong structural constraints for modeling whisker-trigeminal processing at biological scales, constraints which smaller, mouse-relevant object sets might fail to impose. Paralleling findings in vision~\citep{yamins2014performance,khaligh2014deep}, where training on large-scale object categorization leads to generalizable representations beyond specific categories, we similarly suggest that ShapeNet's diversity provides meaningful constraints on network structure irrespective of exact object identity. 
Additionally, by demonstrating that self-supervised learning (SSL) methods yield neural representations comparable (if not marginally better) to supervised approaches, we directly address an open direction highlighted by~\citet{zhuang2017toward}, who emphasized the need for developing more ethologically relevant yet practically scalable tasks for studying tactile processing.

In fact, we found specifically that \emph{contrastive} SSL loss functions consistently reached neural alignment scores similar to the best supervised models, confirming its utility as an ethologically-relevant but label-free approach to tactile representation learning (Fig.~\ref{fig:neural_eval_ssl}), and outperforming other \emph{non-contrastive} self-supervised approaches such as autoencoding.
The SimCLR-optimized IntersectionRNN EAD achieved the best absolute neural alignment score across all models overall, and was comparable in its neural alignment to its supervised (and more parameter-dense) variant (rightmost green bars in Fig.~\ref{fig:neural_fits_doublebar}).
This result mirrors findings in primate visual cortex~\citep{Zhuang2021}, where contrastive SSL methods equally predicted neural response patterns compared to supervised alternatives, suggesting supervised learning serves as a proxy for this more ethologically-relevant loss function.
Interestingly, the parity between contrastive SSL and supervised models observed here differs from mouse visual cortex, where contrastive SSL methods substantially exceeded supervised methods in neural predictivity~\citep{nayebi2021unsupervised}.
Notably, the SSL methods, which are best aligned with tactile neural representations overall, achieve only moderate linear probe categorization performance (Fig.~\ref{fig:neural_eval_ssl}), implying that somatosensory cortex representations might prioritize broader, task-agnostic sensory encoding rather than purely specialized, categorization-driven features.
Furthermore, just as we found for downstream task performance in Fig.~\ref{fig:task_performance}, we also observed it was critical to have tactile-specific SSL augmentations for developing biologically accurate tactile representations, compared to both standard image-based SimCLR augmentations and randomly initialized architecture-fixed controls (Fig.~\ref{fig:neural_eval_augmentations}). We further provide evidence for the design of our tactile augmentation in Appendix~\ref{sec:Additional}.

\section{Discussion}
\label{sec:discuss}
\textbf{Implications for Somatosensory Cortical Processing.}
We developed a novel Encoder-Attender-Decoder (EAD) parameterization of the space of temporal neural network models (TNN) trained to perform tactile recognition on biomechanically realistic force and torque sequences, greatly extending and answering the open question posed by~\citet{zhuang2017toward} of characterizing the ethologically relevant constraints of rodent whisker-based tactile computations.
Our results establish ConvRNN encoders, particularly the IntersectionRNN, as currently superior in tactile categorization performance and neural representational alignment compared to feedforward (ResNet) and state-space models (SSM), suggesting recurrent processing is more relevant overall to the rodent somatosensory system.
Furthermore, we demonstrate that contrastive self-supervised learning (SimCLR), particularly when trained with tactile-specific augmentations, yields neural alignments comparable to supervised methods, highlighting supervised training as a proxy for a more ethologically relevant, label-free representation. 
The modest yet consistent benefits observed from GPT-based Attenders indicate potential attention-like mechanisms operating within hierarchical tactile processing pathways, suggesting a fruitful avenue for experimental validation.

Taken together, our findings indicate that nonlinear recurrent processing plays an essential role in the rodent somatosensory cortex, reflecting neural encoding strategies that prioritize broad, general-purpose tactile representations. 
This work provides the first quantitative characterization of the inductive biases required for tactile algorithms to match brain processing, opening new opportunities for deeper insights into sensory representation learning and somatosensory cortical dynamics.

\textbf{Implications for Embodied AI and Robotics.}
Current artificial tactile sensors and models fall short of animal-like capabilities, limiting the functional use of robots in real-world, unstructured scenarios. 
Our results underscore the importance of biomechanically realistic inputs and temporally recurrent architectures for developing tactile perception models that perceive touch similarly as animals do. 
Our demonstration that tactile-specific (sequential force/torque array) SSL augmentations significantly enhance performance underscores this point, emphasizing the necessity of tailored training methods for robotic tactile systems. 
Future work in embodied robotic systems that leverage these insights could potentially overcome existing sensor limitations, including scaling complexity, stimulus discrimination challenges, and restricted sensing surfaces, thus achieving more robust and nuanced environmental interactions similar to biological organisms.

\textbf{Limitations and Future Directions.}
Our findings do not represent the final word, but rather the beginning of an improved understanding of tactile sensory processing in both animals and machines.
Most importantly, currently available tactile neural datasets remain severely limited in stimulus diversity with the number of object conditions tested, restricting the captured neural variability, and could be the reason why inter-animal consistency values are lower for the statistical average between animals than the current pointwise empirical maximum of 1.3 (Fig.~\ref{fig:neural_fits_doublebar}), indicated by the NeuroAI Turing Test. 
Expanding neural datasets to include broader sets of tactile stimuli and additional animals will thus be crucial for future work, enabling models to approach and potentially surpass this theoretical ceiling of neural predictivity. 

In the longer term, incorporating multimodal sensory integration will be important--particularly combining tactile inputs with other modalities, such as proprioception or vision, through biologically-inspired fusion operations that integrate signals extensively across intermediate layers, as is observed in the brain~\citep{navarro2023visuo}, rather than relying solely on last-layer concatenation, as is commonly done now~\citep{yang2024binding}.
Exploring precisely \textit{where} and \textit{how} these multimodal signals are best fused, using diverse fusion operations (e.g., concatenation, attention, routing), is already supported by our \pytorchtnn library and will yield stronger constraints on models, potentially enabling more robust tactile-driven decision-making in genuinely unstructured scenarios as well as animals do.

\textbf{Broader Impacts.}
Improving robotic tactile sensing to not only match human \emph{capabilities}, but also quantitatively validating, as we explicitly do, shared processing of object properties with human collaborators at the level of \emph{internal} representations, could enhance human safety and efficiency in environments such as healthcare, assistive technologies, and manufacturing in the future.
However, advancements in tactile-enabled robotics might also disrupt employment in sectors relying heavily on manual labor. 
It will therefore be important to thoughtfully manage these technological transitions through proactive policy-making and workforce training.

\newpage

\section*{Acknowledgements}
We thank Katerina Fragkiadaki, Albert Gu, and Roberta ``Bobby'' Klatzky for helpful discussions.
A.N. thanks the Burroughs Wellcome Fund and Google Robotics Award for funding.

\small
\bibliographystyle{abbrvnat}
\bibliography{references}

\begin{thebibliography}{72}
\providecommand{\natexlab}[1]{#1}
\providecommand{\url}[1]{\texttt{#1}}
\expandafter\ifx\csname urlstyle\endcsname\relax
  \providecommand{\doi}[1]{doi: #1}\else
  \providecommand{\doi}{doi: \begingroup \urlstyle{rm}\Url}\fi

\bibitem[Ahl(1986)]{ahl1986role}
A.~S. Ahl.
\newblock The role of vibrissae in behavior: a status review.
\newblock \emph{Veterinary research communications}, 10\penalty0 (1):\penalty0 245--268, 1986.

\bibitem[Armstrong-James et~al.(1992)Armstrong-James, Fox, and Das-Gupta]{armstrong1992flow}
M.~Armstrong-James, K.~Fox, and A.~Das-Gupta.
\newblock Flow of excitation within rat barrel cortex on striking a single vibrissa.
\newblock \emph{Journal of neurophysiology}, 68\penalty0 (4):\penalty0 1345--1358, 1992.

\bibitem[Assaf et~al.(2016)Assaf, Wilson, Anderson, Dean, Porrill, and Pearson]{assaf2016visual}
T.~Assaf, E.~D. Wilson, S.~Anderson, P.~Dean, J.~Porrill, and M.~J. Pearson.
\newblock Visual-tactile sensory map calibration of a biomimetic whiskered robot.
\newblock In \emph{2016 IEEE International Conference on Robotics and Automation (ICRA)}, pages 967--972. IEEE, 2016.

\bibitem[Ba et~al.(2016)Ba, Kiros, and Hinton]{ba2016layernormalization}
J.~L. Ba, J.~R. Kiros, and G.~E. Hinton.
\newblock Layer normalization, 2016.
\newblock URL \url{https://arxiv.org/abs/1607.06450}.

\bibitem[Bakhtiari et~al.(2021)Bakhtiari, Mineault, Lillicrap, Pack, and Richards]{bakhtiari2021functional}
S.~Bakhtiari, P.~Mineault, T.~Lillicrap, C.~Pack, and B.~Richards.
\newblock The functional specialization of visual cortex emerges from training parallel pathways with self-supervised predictive learning.
\newblock \emph{Advances in Neural Information Processing Systems}, 34:\penalty0 25164--25178, 2021.

\bibitem[Belli et~al.(2017)Belli, Yang, Bresee, and Hartmann]{belli2017variations}
H.~M. Belli, A.~E. Yang, C.~S. Bresee, and M.~J. Hartmann.
\newblock Variations in vibrissal geometry across the rat mystacial pad: base diameter, medulla, and taper.
\newblock \emph{Journal of Neurophysiology}, 117\penalty0 (4):\penalty0 1807--1820, 2017.

\bibitem[Belli et~al.(2018)Belli, Bresee, Graff, and Hartmann]{belli2018quantifying}
H.~M. Belli, C.~S. Bresee, M.~M. Graff, and M.~J. Hartmann.
\newblock Quantifying the three-dimensional facial morphology of the laboratory rat with a focus on the vibrissae.
\newblock \emph{PLoS One}, 13\penalty0 (4):\penalty0 e0194981, 2018.

\bibitem[Bhirangi et~al.(2021)Bhirangi, Hellebrekers, Majidi, and Gupta]{bhirangi2021reskin}
R.~Bhirangi, T.~Hellebrekers, C.~Majidi, and A.~Gupta.
\newblock Reskin: versatile, replaceable, lasting tactile skins.
\newblock \emph{arXiv preprint arXiv:2111.00071}, 2021.

\bibitem[Bhirangi et~al.(2024{\natexlab{a}})Bhirangi, Pattabiraman, Erciyes, Cao, Hellebrekers, and Pinto]{bhirangi2024anyskin}
R.~Bhirangi, V.~Pattabiraman, E.~Erciyes, Y.~Cao, T.~Hellebrekers, and L.~Pinto.
\newblock Anyskin: Plug-and-play skin sensing for robotic touch.
\newblock \emph{arXiv preprint arXiv:2409.08276}, 2024{\natexlab{a}}.

\bibitem[Bhirangi et~al.(2024{\natexlab{b}})Bhirangi, Wang, Pattabiraman, Majidi, Gupta, Hellebrekers, and Pinto]{bhirangi2024hierarchical}
R.~Bhirangi, C.~Wang, V.~Pattabiraman, C.~Majidi, A.~Gupta, T.~Hellebrekers, and L.~Pinto.
\newblock Hierarchical state space models for continuous sequence-to-sequence modeling.
\newblock \emph{arXiv preprint arXiv:2402.10211}, 2024{\natexlab{b}}.

\bibitem[Bosman et~al.(2011)Bosman, Houweling, Owens, Tanke, Shevchouk, Rahmati, Teunissen, Ju, Gong, Koekkoek, et~al.]{bosman2011anatomical}
L.~W. Bosman, A.~R. Houweling, C.~B. Owens, N.~Tanke, O.~T. Shevchouk, N.~Rahmati, W.~H. Teunissen, C.~Ju, W.~Gong, S.~K. Koekkoek, et~al.
\newblock Anatomical pathways involved in generating and sensing rhythmic whisker movements.
\newblock \emph{Frontiers in integrative neuroscience}, 5:\penalty0 53, 2011.

\bibitem[Bottou(2010)]{bottou2010large}
L.~Bottou.
\newblock Large-scale machine learning with stochastic gradient descent.
\newblock In \emph{Proceedings of COMPSTAT'2010: 19th International Conference on Computational StatisticsParis France, August 22-27, 2010 Keynote, Invited and Contributed Papers}, pages 177--186. Springer, 2010.

\bibitem[Bresee et~al.(2023)Bresee, Belli, Luo, and Hartmann]{bresee_mice_morphology}
C.~S. Bresee, H.~M. Belli, Y.~Luo, and M.~J.~Z. Hartmann.
\newblock Comparative morphology of the whiskers and faces of mice (mus musculus) and rats (rattus norvegicus).
\newblock \emph{Journal of Experimental Biology}, 226\penalty0 (19):\penalty0 jeb245597, 10 2023.
\newblock ISSN 0022-0949.
\newblock \doi{10.1242/jeb.245597}.
\newblock URL \url{https://doi.org/10.1242/jeb.245597}.

\bibitem[Cao and Yamins(2024)]{cao2024explanatory2}
R.~Cao and D.~Yamins.
\newblock Explanatory models in neuroscience, part 2: Functional intelligibility and the contravariance principle.
\newblock \emph{Cognitive Systems Research}, 85:\penalty0 101200, 2024.

\bibitem[Chang et~al.(2015)Chang, Funkhouser, Guibas, Hanrahan, Huang, Li, Savarese, Savva, Song, Su, et~al.]{chang2015shapenet}
A.~X. Chang, T.~Funkhouser, L.~Guibas, P.~Hanrahan, Q.~Huang, Z.~Li, S.~Savarese, M.~Savva, S.~Song, H.~Su, et~al.
\newblock Shapenet: An information-rich 3d model repository.
\newblock \emph{arXiv preprint arXiv:1512.03012}, 2015.

\bibitem[Chen et~al.(2020)Chen, Kornblith, Norouzi, and Hinton]{chen2020simple}
T.~Chen, S.~Kornblith, M.~Norouzi, and G.~Hinton.
\newblock A simple framework for contrastive learning of visual representations.
\newblock In \emph{International conference on machine learning}, pages 1597--1607. PmLR, 2020.

\bibitem[Chen and He(2021)]{Chen2020Siam}
X.~Chen and K.~He.
\newblock Exploring simple siamese representation learning.
\newblock In \emph{Proceedings of the IEEE/CVF conference on computer vision and pattern recognition}, pages 15750--15758, 2021.

\bibitem[Cheung et~al.(2019)Cheung, Maire, Kim, Sy, and Hires]{cheung2019sensorimotor}
J.~Cheung, P.~Maire, J.~Kim, J.~Sy, and S.~A. Hires.
\newblock The sensorimotor basis of whisker-guided anteroposterior object localization in head-fixed mice.
\newblock \emph{Current Biology}, 29\penalty0 (18):\penalty0 3029--3040, 2019.

\bibitem[Cho et~al.(2014)Cho, van Merrienboer, Gulcehre, Bahdanau, Bougares, Schwenk, and Bengio]{cho2014learningphraserepresentationsusing}
K.~Cho, B.~van Merrienboer, C.~Gulcehre, D.~Bahdanau, F.~Bougares, H.~Schwenk, and Y.~Bengio.
\newblock Learning phrase representations using rnn encoder-decoder for statistical machine translation, 2014.
\newblock URL \url{https://arxiv.org/abs/1406.1078}.

\bibitem[Collins et~al.(2017)Collins, Sohl-Dickstein, and Sussillo]{Collins2017}
J.~Collins, J.~Sohl-Dickstein, and D.~Sussillo.
\newblock Capacity and trainability in recurrent neural networks.
\newblock In \emph{ICLR}, 2017.

\bibitem[Coumans and Bai(2016--2021)]{bullet_coumans2021}
E.~Coumans and Y.~Bai.
\newblock Pybullet, a python module for physics simulation for games, robotics and machine learning.
\newblock \url{http://pybullet.org}, 2016--2021.

\bibitem[Dahiya et~al.(2009)Dahiya, Metta, Valle, and Sandini]{dahiya2009tactile}
R.~S. Dahiya, G.~Metta, M.~Valle, and G.~Sandini.
\newblock Tactile sensing—from humans to humanoids.
\newblock \emph{IEEE transactions on robotics}, 26\penalty0 (1):\penalty0 1--20, 2009.

\bibitem[Dehnhardt et~al.(2001)Dehnhardt, Mauck, Hanke, and Bleckmann]{dehnhardt2001hydrodynamic}
G.~Dehnhardt, B.~Mauck, W.~Hanke, and H.~Bleckmann.
\newblock Hydrodynamic trail-following in harbor seals (phoca vitulina).
\newblock \emph{Science}, 293\penalty0 (5527):\penalty0 102--104, 2001.

\bibitem[Feather et~al.(2019)Feather, Durango, Gonzalez, and McDermott]{feather2019metamers}
J.~Feather, A.~Durango, R.~Gonzalez, and J.~McDermott.
\newblock Metamers of neural networks reveal divergence from human perceptual systems.
\newblock In \emph{Advances in Neural Information Processing Systems}, pages 10078--10089, 2019.

\bibitem[Feather* et~al.(2025)Feather*, Khosla*, Murty*, and Nayebi*]{feather2025brain}
J.~Feather*, M.~Khosla*, N.~Murty*, and A.~Nayebi*.
\newblock Brain-model evaluations need the neuroai turing test.
\newblock \emph{arXiv preprint arXiv:2502.16238}, 2025.

\bibitem[Felleman and Van~Essen(1991)]{felleman1991distributed}
D.~J. Felleman and D.~C. Van~Essen.
\newblock Distributed hierarchical processing in the primate cerebral cortex.
\newblock \emph{Cerebral cortex (New York, NY: 1991)}, 1\penalty0 (1):\penalty0 1--47, 1991.

\bibitem[Freeman and Lemen(2008)]{freeman2008simple}
P.~W. Freeman and C.~A. Lemen.
\newblock A simple morphological predictor of bite force in rodents.
\newblock \emph{Journal of Zoology}, 275\penalty0 (4):\penalty0 418--422, 2008.

\bibitem[Grant(2025)]{grant2025whisker}
R.~A. Grant.
\newblock Can we study whisker movements to gain insights into the natural sensory behaviours of mammals?
\newblock \emph{The Journal of Physiology}, 2025.

\bibitem[Gu and Dao(2023)]{mamba}
A.~Gu and T.~Dao.
\newblock Mamba: Linear-time sequence modeling with selective state spaces.
\newblock \emph{arXiv preprint arXiv:2312.00752}, 2023.

\bibitem[Gu et~al.(2021)Gu, Goel, and R{\'e}]{gu2021efficiently}
A.~Gu, K.~Goel, and C.~R{\'e}.
\newblock Efficiently modeling long sequences with structured state spaces.
\newblock \emph{arXiv preprint arXiv:2111.00396}, 2021.

\bibitem[Haldar et~al.(2024)Haldar, Peng, and Pinto]{haldar2024baku}
S.~Haldar, Z.~Peng, and L.~Pinto.
\newblock Baku: An efficient transformer for multi-task policy learning.
\newblock \emph{arXiv preprint arXiv:2406.07539}, 2024.

\bibitem[He et~al.(2016)He, Zhang, Ren, and Sun]{He2016}
K.~He, X.~Zhang, S.~Ren, and J.~Sun.
\newblock Deep residual learning for image recognition.
\newblock In \emph{Proceedings of the IEEE conference on computer vision and pattern recognition}, pages 770--778, 2016.

\bibitem[Hires et~al.(2015)Hires, Gutnisky, Yu, O'Connor, and Svoboda]{andrew2015low}
S.~A. Hires, D.~A. Gutnisky, J.~Yu, D.~H. O'Connor, and K.~Svoboda.
\newblock Low-noise encoding of active touch by layer 4 in the somatosensory cortex.
\newblock \emph{elife}, 4:\penalty0 e06619, 2015.

\bibitem[Hochreiter and Schmidhuber(1997)]{Hochreiter1997}
S.~Hochreiter and J.~Schmidhuber.
\newblock Long short-term memory.
\newblock \emph{Neural Computation}, 9\penalty0 (8):\penalty0 1735--1780, 1997.

\bibitem[Kell* et~al.(2018)Kell*, Yamins*, Shook, Norman-Haignere, and McDermott]{kell2018task}
A.~J. Kell*, D.~L. Yamins*, E.~N. Shook, S.~V. Norman-Haignere, and J.~H. McDermott.
\newblock A task-optimized neural network replicates human auditory behavior, predicts brain responses, and reveals a cortical processing hierarchy.
\newblock \emph{Neuron}, 98\penalty0 (3):\penalty0 630--644, 2018.

\bibitem[Kent and Bergbreiter(2024)]{kent2024flow}
T.~A. Kent and S.~Bergbreiter.
\newblock Flow shadowing: A method to detect multiple flow headings using an array of densely packed whisker-inspired sensors.
\newblock In \emph{2024 IEEE International Conference on Robotics and Automation (ICRA)}, pages 17843--17849. IEEE, 2024.

\bibitem[Kent et~al.(2023)Kent, Emnett, Babaei, Hartmann, and Bergbreiter]{kent2023identifying}
T.~A. Kent, H.~Emnett, M.~Babaei, M.~J. Hartmann, and S.~Bergbreiter.
\newblock Identifying contact distance uncertainty in whisker sensing with tapered, flexible whiskers.
\newblock In \emph{2023 IEEE International Conference on Robotics and Automation (ICRA)}, pages 607--613. IEEE, 2023.

\bibitem[Kerr et~al.(2007)Kerr, De~Kock, Greenberg, Bruno, Sakmann, and Helmchen]{kerr2007spatial}
J.~N. Kerr, C.~P. De~Kock, D.~S. Greenberg, R.~M. Bruno, B.~Sakmann, and F.~Helmchen.
\newblock Spatial organization of neuronal population responses in layer 2/3 of rat barrel cortex.
\newblock \emph{Journal of neuroscience}, 27\penalty0 (48):\penalty0 13316--13328, 2007.

\bibitem[Khaligh-Razavi and Kriegeskorte(2014)]{khaligh2014deep}
S.-M. Khaligh-Razavi and N.~Kriegeskorte.
\newblock Deep supervised, but not unsupervised, models may explain it cortical representation.
\newblock \emph{PLoS computational biology}, 10\penalty0 (11):\penalty0 e1003915, 2014.

\bibitem[Kingma and Ba(2015)]{kingma2014adam}
D.~P. Kingma and J.~Ba.
\newblock Adam: A method for stochastic optimization.
\newblock In \emph{ICLR}, 2015.

\bibitem[Knutsen et~al.(2008)Knutsen, Biess, and Ahissar]{knutsen2008vibrissal}
P.~M. Knutsen, A.~Biess, and E.~Ahissar.
\newblock Vibrissal kinematics in 3d: tight coupling of azimuth, elevation, and torsion across different whisking modes.
\newblock \emph{Neuron}, 59\penalty0 (1):\penalty0 35--42, 2008.

\bibitem[Kriegeskorte et~al.(2008)Kriegeskorte, Mur, and Bandettini]{kriegeskorte2008representational}
N.~Kriegeskorte, M.~Mur, and P.~A. Bandettini.
\newblock Representational similarity analysis-connecting the branches of systems neuroscience.
\newblock \emph{Frontiers in systems neuroscience}, 2:\penalty0 249, 2008.

\bibitem[Lederman and Klatzky(2009)]{lederman2009haptic}
S.~J. Lederman and R.~L. Klatzky.
\newblock Haptic perception: A tutorial.
\newblock \emph{Attention, Perception, \& Psychophysics}, 71\penalty0 (7):\penalty0 1439--1459, 2009.

\bibitem[Loshchilov and Hutter(2019)]{loshchilov2019decoupledweightdecayregularization}
I.~Loshchilov and F.~Hutter.
\newblock Decoupled weight decay regularization, 2019.
\newblock URL \url{https://arxiv.org/abs/1711.05101}.

\bibitem[Mamou and Ghorbel(2009)]{vhacd2009}
K.~Mamou and F.~Ghorbel.
\newblock A simple and efficient approach for 3d mesh approximate convex decomposition.
\newblock pages 3501--3504, 11 2009.
\newblock \doi{10.1109/ICIP.2009.5414068}.

\bibitem[Michaels et~al.()Michaels, Schaffelhofer, Agudelo-Toro, and Scherberger]{michaels2020goal}
J.~A. Michaels, S.~Schaffelhofer, A.~Agudelo-Toro, and H.~Scherberger.
\newblock A goal-driven modular neural network predicts parietofrontal neural dynamics during grasping.
\newblock 117\penalty0 (50):\penalty0 32124--32135.

\bibitem[Moore et~al.(2015)Moore, Mercer~Lindsay, Desch{\^e}nes, and Kleinfeld]{moore2015vibrissa}
J.~D. Moore, N.~Mercer~Lindsay, M.~Desch{\^e}nes, and D.~Kleinfeld.
\newblock Vibrissa self-motion and touch are reliably encoded along the same somatosensory pathway from brainstem through thalamus.
\newblock \emph{PLoS biology}, 13\penalty0 (9):\penalty0 e1002253, 2015.

\bibitem[Navarro-Guerrero et~al.(2023)Navarro-Guerrero, Toprak, Josifovski, and Jamone]{navarro2023visuo}
N.~Navarro-Guerrero, S.~Toprak, J.~Josifovski, and L.~Jamone.
\newblock Visuo-haptic object perception for robots: an overview.
\newblock \emph{Autonomous Robots}, 47\penalty0 (4):\penalty0 377--403, 2023.

\bibitem[Nayebi* et~al.(2018)Nayebi*, Bear*, Kubilius*, Kar, Ganguli, Sussillo, DiCarlo, and Yamins]{nayebi2018task}
A.~Nayebi*, D.~Bear*, J.~Kubilius*, K.~Kar, S.~Ganguli, D.~Sussillo, J.~J. DiCarlo, and D.~L. Yamins.
\newblock Task-driven convolutional recurrent models of the visual system.
\newblock In S.~Bengio, H.~Wallach, H.~Larochelle, K.~Grauman, N.~Cesa-Bianchi, and R.~Garnett, editors, \emph{Advances in Neural Information Processing Systems}, volume~31. Curran Associates, Inc., 2018.

\bibitem[Nayebi et~al.(2021)Nayebi, Attinger, Campbell, Hardcastle, Low, Mallory, Mel, Sorscher, Williams, Ganguli, Giocomo, and Yamins]{nayebi2021explaining}
A.~Nayebi, A.~Attinger, M.~Campbell, K.~Hardcastle, I.~Low, C.~Mallory, G.~Mel, B.~Sorscher, A.~Williams, S.~Ganguli, L.~M. Giocomo, and D.~L. Yamins.
\newblock Explaining heterogeneity in medial entorhinal cortex with task-driven neural networks.
\newblock \emph{Advances in Neural Information Processing Systems}, 34, 2021.

\bibitem[Nayebi et~al.(2022)Nayebi, Sagastuy-Brena, Bear, Kar, Kubilius, Ganguli, Sussillo, DiCarlo, and Yamins]{nayebi2022}
A.~Nayebi, J.~Sagastuy-Brena, D.~M. Bear, K.~Kar, J.~Kubilius, S.~Ganguli, D.~Sussillo, J.~J. DiCarlo, and D.~L. Yamins.
\newblock Recurrent connections in the primate ventral visual stream mediate a tradeoff between task performance and network size during core object recognition.
\newblock \emph{Neural Computation}, 34:\penalty0 1652--1675, 2022.

\bibitem[Nayebi* et~al.(2023)Nayebi*, Kong*, Zhuang, Gardner, Norcia, and Yamins]{nayebi2021unsupervised}
A.~Nayebi*, N.~C. Kong*, C.~Zhuang, J.~L. Gardner, A.~M. Norcia, and D.~L. Yamins.
\newblock Mouse visual cortex as a limited resource system that self-learns an ecologically-general representation.
\newblock \emph{PLOS Computational Biology}, 19, 2023.

\bibitem[Olshausen and Field(1996)]{olshausen1996emergence}
B.~A. Olshausen and D.~J. Field.
\newblock Emergence of simple-cell receptive field properties by learning a sparse code for natural images.
\newblock \emph{Nature}, 381\penalty0 (6583):\penalty0 607--609, 1996.

\bibitem[Pearson et~al.(2011)Pearson, Mitchinson, Sullivan, Pipe, and Prescott]{pearson2011biomimetic}
M.~J. Pearson, B.~Mitchinson, J.~C. Sullivan, A.~G. Pipe, and T.~J. Prescott.
\newblock Biomimetic vibrissal sensing for robots.
\newblock \emph{Philosophical Transactions of the Royal Society B: Biological Sciences}, 366\penalty0 (1581):\penalty0 3085--3096, 2011.

\bibitem[Rodgers(2022)]{rodgers2022detailed}
C.~C. Rodgers.
\newblock A detailed behavioral, videographic, and neural dataset on object recognition in mice.
\newblock \emph{Scientific Data}, 9\penalty0 (1):\penalty0 620, 2022.

\bibitem[Schrimpf et~al.(2020)Schrimpf, Kubilius, Lee, Murty, Ajemian, and DiCarlo]{schrimpf2020integrative}
M.~Schrimpf, J.~Kubilius, M.~J. Lee, N.~A.~R. Murty, R.~Ajemian, and J.~J. DiCarlo.
\newblock Integrative benchmarking to advance neurally mechanistic models of human intelligence.
\newblock \emph{Neuron}, 108\penalty0 (3):\penalty0 413--423, 2020.

\bibitem[Schrimpf et~al.(2021)Schrimpf, Blank, Tuckute, Kauf, Hosseini, Kanwisher, Tenenbaum, and Fedorenko]{schrimpf2021neural}
M.~Schrimpf, I.~A. Blank, G.~Tuckute, C.~Kauf, E.~A. Hosseini, N.~Kanwisher, J.~B. Tenenbaum, and E.~Fedorenko.
\newblock The neural architecture of language: Integrative modeling converges on predictive processing.
\newblock \emph{Proceedings of the National Academy of Sciences}, 118\penalty0 (45):\penalty0 e2105646118, 2021.

\bibitem[Shimonomura(2019)]{shimonomura2019tactile}
K.~Shimonomura.
\newblock Tactile image sensors employing camera: A review.
\newblock \emph{Sensors}, 19\penalty0 (18):\penalty0 3933, 2019.

\bibitem[Simon et~al.(2023)Simon, Ren, Piqu{\'e}, Snyder, Barretto, Hultmark, and Majumdar]{simon2023flowdrone}
N.~Simon, A.~Z. Ren, A.~Piqu{\'e}, D.~Snyder, D.~Barretto, M.~Hultmark, and A.~Majumdar.
\newblock Flowdrone: wind estimation and gust rejection on uavs using fast-response hot-wire flow sensors.
\newblock In \emph{2023 IEEE International Conference on Robotics and Automation (ICRA)}, pages 5393--5399. IEEE, 2023.

\bibitem[Sofroniew and Svoboda(2015)]{sofroniew2015whisking}
N.~J. Sofroniew and K.~Svoboda.
\newblock Whisking.
\newblock \emph{Current Biology}, 25\penalty0 (4):\penalty0 R137--R140, 2015.

\bibitem[Spoerer et~al.(2017)Spoerer, McClure, and Kriegeskorte]{Spoerer2017}
C.~J. Spoerer, P.~McClure, and N.~Kriegeskorte.
\newblock Recurrent convolutional neural networks: a better model of biological object recognition.
\newblock \emph{Front. Psychol.}, 8:\penalty0 1--14, 2017.

\bibitem[Staiger and Petersen(2021)]{staiger2021neuronal}
J.~F. Staiger and C.~C. Petersen.
\newblock Neuronal circuits in barrel cortex for whisker sensory perception.
\newblock \emph{Physiological reviews}, 101\penalty0 (1):\penalty0 353--415, 2021.

\bibitem[Sterbing-D'Angelo et~al.(2011)Sterbing-D'Angelo, Chadha, Chiu, Falk, Xian, Barcelo, Zook, and Moss]{sterbing2011bat}
S.~Sterbing-D'Angelo, M.~Chadha, C.~Chiu, B.~Falk, W.~Xian, J.~Barcelo, J.~M. Zook, and C.~F. Moss.
\newblock Bat wing sensors support flight control.
\newblock \emph{Proceedings of the National Academy of Sciences}, 108\penalty0 (27):\penalty0 11291--11296, 2011.

\bibitem[Sussillo et~al.()Sussillo, Churchland, Kaufman, and Shenoy]{sussillo2015neural}
D.~Sussillo, M.~M. Churchland, M.~T. Kaufman, and K.~V. Shenoy.
\newblock A neural network that finds a naturalistic solution for the production of muscle activity.
\newblock 18\penalty0 (7):\penalty0 1025--1033.

\bibitem[Ward-Cherrier et~al.(2018)Ward-Cherrier, Pestell, Cramphorn, Winstone, Giannaccini, Rossiter, and Lepora]{ward2018tactip}
B.~Ward-Cherrier, N.~Pestell, L.~Cramphorn, B.~Winstone, M.~E. Giannaccini, J.~Rossiter, and N.~F. Lepora.
\newblock The tactip family: Soft optical tactile sensors with 3d-printed biomimetic morphologies.
\newblock \emph{Soft robotics}, 5\penalty0 (2):\penalty0 216--227, 2018.

\bibitem[Yamins et~al.()Yamins, Hong, Cadieu, Solomon, Seibert, and DiCarlo]{yamins2014performance}
D.~L. Yamins, H.~Hong, C.~F. Cadieu, E.~A. Solomon, D.~Seibert, and J.~J. DiCarlo.
\newblock Performance-optimized hierarchical models predict neural responses in higher visual cortex.
\newblock 111\penalty0 (23):\penalty0 8619--8624.

\bibitem[Yang et~al.(2024)Yang, Feng, Chen, Park, Wang, Dou, Zeng, Chen, Gangopadhyay, Owens, et~al.]{yang2024binding}
F.~Yang, C.~Feng, Z.~Chen, H.~Park, D.~Wang, Y.~Dou, Z.~Zeng, X.~Chen, R.~Gangopadhyay, A.~Owens, et~al.
\newblock Binding touch to everything: Learning unified multimodal tactile representations.
\newblock In \emph{Proceedings of the IEEE/CVF Conference on Computer Vision and Pattern Recognition}, pages 26340--26353, 2024.

\bibitem[You et~al.()You, Gitman, and Ginsburg]{you2017large}
Y.~You, I.~Gitman, and B.~Ginsburg.
\newblock Large batch training of convolutional networks.

\bibitem[Yuan et~al.(2017)Yuan, Dong, and Adelson]{yuan2017gelsight}
W.~Yuan, S.~Dong, and E.~H. Adelson.
\newblock Gelsight: High-resolution robot tactile sensors for estimating geometry and force.
\newblock \emph{Sensors}, 17\penalty0 (12):\penalty0 2762, 2017.

\bibitem[Zhuang et~al.(2017)Zhuang, Kubilius, Hartmann, and Yamins]{zhuang2017toward}
C.~Zhuang, J.~Kubilius, M.~J. Hartmann, and D.~L. Yamins.
\newblock Toward goal-driven neural network models for the rodent whisker-trigeminal system.
\newblock \emph{Advances in Neural Information Processing Systems}, 30, 2017.

\bibitem[Zhuang et~al.(2021)Zhuang, Yan, Nayebi, Schrimpf, Frank, DiCarlo, and Yamins]{Zhuang2021}
C.~Zhuang, S.~Yan, A.~Nayebi, M.~Schrimpf, M.~C. Frank, J.~J. DiCarlo, and D.~L. Yamins.
\newblock Unsupervised neural network models of the ventral visual stream.
\newblock \emph{Proceedings of the National Academy of Sciences}, 118\penalty0 (3), 2021.

\bibitem[Zweifel et~al.(2021)Zweifel, Bush, Abraham, Murphey, and Hartmann]{zweifel2021dynamical}
N.~O. Zweifel, N.~E. Bush, I.~Abraham, T.~D. Murphey, and M.~J. Hartmann.
\newblock A dynamical model for generating synthetic data to quantify active tactile sensing behavior in the rat.
\newblock \emph{Proceedings of the National Academy of Sciences}, 118\penalty0 (27):\penalty0 e2011905118, 2021.

\end{thebibliography}


\newpage
\appendix
\setcounter{figure}{0}
\renewcommand{\thefigure}{A\arabic{figure}}
\setcounter{section}{0}
\renewcommand{\thesection}{A\arabic{section}}
\normalsize
\section*{Appendix}

\section*{Table of Contents}

We detail the contents in the appendix below.
\begin{itemize}
    \item[A1.] \hyperref[sec:Dataset]{\textbf{Whisk Dataset Generation}} details the parameters and the modifications of the simulator for generating our whisking dataset for training. 
    \item[A2.] \hyperref[sec:Model]{\textbf{Model Definitions}} present the mathematical definitions of the RNNs used in the ConvRNNs (i.e, Zhuang's model~\citep{zhuang2017toward}) in our experiments.
    \item[A3.] \hyperref[sec:Model]{\textbf{Model Training}} includes the optimizer, scheduler, and learning rate configurations for training different models for both supervised and self-supervised learning.
    \item[A4.] \hyperref[sec:Neural]{\textbf{Neural Evaluation}} explains how the neural fit scores are calculated, and presents visualization of model parameters vs. neural fit, per layer neural fit scores, and representational dissimilarity matrices for neural evaluation. 
    \item[A5.] \hyperref[sec:Additional]{\textbf{Additional Experiments}} show results on stimuli decoding based on learned representations and self-supervised learning performance with temporal masking as the augmentation.
\end{itemize}

\vspace{0.5cm}

\section{Whisk Dataset Generation}
\label{sec:Dataset}

Our whisking dataset uses the same 9981 ShapeNet objects and 117 category labels as in \citet{zhuang2017toward}, but using an improved whisker model and various sweep augmentations, which are listed in Table~\ref{tab:sweeps}. 

\begin{table}[h]
\centering
\begin{tabular}{cc|ccccc|c}
\toprule
& Sim. Freq. & Speed & Height & Rotation & Distance & Size & Total Sweeps \\
\midrule
(1) & 1000 Hz & 30 mm/s & -5, 0 mm & 0, 30, 90, 120° & 5, 8 mm & 40 mm & 316,192 \\
(2) & 110 Hz & [30$\sim$60] & -3, 0, 3 & [0$\sim$359] & 5 & [20$\sim$60] & 2,076,048 \\
\bottomrule
\end{tabular}
\vspace{0.05in}
\caption{
\textbf{Sweep Augmentations}
used for the two whisking datasets, which we refer to as (1) Low-Variation High-Fidelity and (2) High-Variation Low-Fidelity.
Simulation Frequency refers to the frequency corresponding to the physics timestep used in Bullet physics engine \citep{bullet_coumans2021}, the backend for WHISKiT simulator \citep{zweifel2021dynamical}.
For dataset (2), the speed, rotation, and size is each sampled from the range 26 times.
The total number of sweeps equals the number of sweep augmentation combinations $\times$ 9981 objects.
}
\label{tab:sweeps}
\end{table}



\textbf{WHISKiT Simulator Modifications}.
We make a few enhancements to the original WHISKiT simulator \citep{zweifel2021dynamical}:
\vspace{-6pt}
\begin{itemize}\itemsep0em 
  \item Number of whisker links (where whiskers are modeled as a chain of springs \citep{zhuang2017toward}) are dynamically adjusted by the length of the whisker instead of a fixed number.
  \item Allow for loading objects with or without convex hull.
    In the whisk datasets, objects are loaded with convex hull (using V-HACD \citep{vhacd2009}) to keep collisions more stable. In generating the 6 simulated stimuli for model input neural evaluation, convex hull was not used in order to preserve the concavity of the ``concave'' object. The object was geometrically simple enough to not affect collision stability.
  \item Add camera settings for viewing in orthographic or perspective projection.
  \item Load multiple mice/rats at a time.
\end{itemize}

All our code for the simulation, model training, and neural data analysis is available on GitHub: \url{https://github.com/neuroagents-lab/2025-tactile-whisking}

\clearpage

\section{Model Definitions}
\label{sec:Model}

We provide the update rules for UGRNN and IntersectionRNN in our experiments as follows. For other RNN variants, please refer to~\cite{nayebi2022}.

\begin{itemize}
    \item $x_t^\ell$: input at time $t$ and layer $\ell$
    \item $s_t^\ell$: state at time $t$ and layer $\ell$
    \item $W^\ell, U^\ell$: learnable weight matrices at layer $\ell$
    \item $b^\ell$: bias terms at layer $\ell$
    \item $\circ$: element-wise multiplication
    \item $\sigma$: sigmoid function
    \item $\tanh$: hyperbolic tangent function
    \item $\mathrm{ReLU}$: rectified linear unit function
    \item $*$: linear transformation or convolution (depending on context)
\end{itemize}

\subsection{UGRNN}
\begin{align}
c_t^\ell &= \tanh\!\left(W_c^\ell * x_t^\ell + U_c^\ell * s_{t-1}^\ell + b_c^\ell \right) \\
g_t^\ell &= \sigma\!\left(W_g^\ell * x_t^\ell + U_g^\ell * s_{t-1}^\ell + b_g^\ell + 1 \right) \\
s_t^\ell &= g_t^\ell \circ s_{t-1}^\ell + \left(1 - g_t^\ell \right) \circ c_t^\ell \\
h_t^\ell &= s_t^\ell
\end{align}
\subsection{IntersectionRNN}
\begin{align}
m_t^\ell &= \tanh\!\left(W_m^\ell * x_t^\ell + U_m^\ell * s_{t-1}^\ell + b_m^\ell \right) \\
n_t^\ell &= \mathrm{ReLU}\!\left(W_n^\ell * x_t^\ell + U_n^\ell * s_{t-1}^\ell + b_n^\ell \right) \\
p_t^\ell &= \sigma\!\left(W_p^\ell * x_t^\ell + U_p^\ell * s_{t-1}^\ell + b_p^\ell + 1 \right) \\
y_t^\ell &= \sigma\!\left(W_y^\ell * x_t^\ell + U_y^\ell * s_{t-1}^\ell + b_y^\ell + 1 \right) \\
s_t^\ell &= p_t^\ell \circ s_{t-1}^\ell + \left(1 - p_t^\ell \right) \circ m_t^\ell \\
h_t^\ell &= y_t^\ell \circ x_t^\ell + \left(1 - y_t^\ell \right) \circ n_t^\ell
\end{align}

\clearpage

\section{Model Training}
\label{sec:Model}
All of our experiments are conducted on NVIDIA A6000 GPUs. For supervised learning, we use a batch size of 256 for all the models and train for 100 epochs. For SSL, during the pre-training stage, we use a batch size of 256 for SimCLR and autoencoding (AE), and a batch size of 1024 for SimSiam, following~\cite{nayebi2021unsupervised}, and train for 100 epochs. During the linear probing stage, we freeze the checkpoint saved with the lowest validation loss and add a trainable linear layer. We further train such a model with labels for 100 epochs, with a batch size of 256, an initial learning rate of 0.1, with the StepLR scheduler, and the SGD optimizer with momentum~\citep{bottou2010large}.

We detail the optimizers~\citep{bottou2010large, kingma2014adam, loshchilov2019decoupledweightdecayregularization, you2017large} and schedulers, and their configurations we used in supervised learning and SSL in Table~\ref{tab:opt_sch_cfg}.

\begin{table}[h]
\centering
\begin{tabular}{c|c}
\toprule
\textbf{Optimizer} & \textbf{Configuration} \\
\midrule
SGD & $\texttt{momentum}=0.9$, $\texttt{weight-decay}=10^{-4}$ \\
Adam & $\texttt{weight-decay}=10^{-4}$ \\
AdamW & $\texttt{weight-decay}=10^{-4}$ \\
LARS & $\texttt{momentum}=0.9$, $\texttt{weight-decay}=10^{-4}$ \\
\midrule
\textbf{Scheduler} & \textbf{Configuration} \\
\midrule
StepLR & $\texttt{step\_size}=30$ \\
ConstantLR & fixed learning rate \\
CosineLR & $\texttt{warmup\_epoch}=10$ \\
CosineAnnealing & $\texttt{min\_lr}=0.0$, $\texttt{warmup\_epoch}=10$, $\texttt{warmup\_ratio}=10^{-4}$ \\
\bottomrule
\end{tabular}
\vspace{0.05in}
\caption{\textbf{Optimizers and Schedulers} with default configurations of supervised learning and SSL when training different model architectures.}
\label{tab:opt_sch_cfg}
\end{table}

For supervised learning, we present the model training configurations in Table~\ref{tab:sup_model_cfg}, where we detail the choices of optimizer, scheduler, learning rate, the encoder and attender of our EAD architecture, and we omit the decoder due to space limits, as it is always a linear layer/MLP. As Zhuang+GPT is the best performing supervised model in terms of classification accuracy, we explore different variants of Zhuang's encoder with attender being GPT.

\begin{table}[h]
\centering
\resizebox{\textwidth}{!}{%
\begin{tabular}{c|c|ccc}
\toprule
\textbf{Encoder} & \textbf{Attender} & \textbf{Optimizer} & \textbf{Scheduler} & \textbf{Learning Rate} \\
\midrule
ResNet & None & SGD & StepLR & $10^{\{-1, -2, -3\}}$ \\
\midrule
\multirow{2}{*}{Zhuang} & \multirow{2}{*}{None} & SGD & StepLR & $10^{\{-1, -2, -3, -4\}}$ \\
   &   & SGD & ConstantLR & $10^{-2}, 5\times10^{-3}$ \\
\midrule
Zhuang-\{UGRNN, & \multirow{2}{*}{None} & \{SGD, Adam, AdamW\} & StepLR & $10^{\{-1, -2, -3, -4\}}$ \\
IntersectionRNN, GRU, LSTM\}                    &                       & (LayerNorm) - AdamW & StepLR & $10^{\{-1, -2, -3, -4\}}$ \\
\midrule
ResNet, Zhuang, Zhuang-\{UGRNN & \multirow{2}{*}{GPT} & AdamW & CosineLR & $10^{-4}$ \\
IntersectionRNN-LN, GRU, LSTM\}, S4                    &                       & \{SGD, AdamW\} & StepLR & $10^{\{-1, -2, -3\}}$ \\
\midrule
\multirow{2}{*}{ResNet, Zhuang, S4} & \multirow{2}{*}{Mamba} & AdamW & CosineLR & $10^{-4}$ \\
                    &                       & \{SGD, AdamW\} & StepLR & $10^{\{-1, -2, -3\}}$ \\
\bottomrule
\end{tabular}
}
\vspace{0.05in}
\caption{
\textbf{Model Training Configurations for Supervised Learning}.
We use \{\} to indicate different choices of a specific component (i.e., encoder, optimizer, learning rate) in the search space. We consider adding layer norm as a variant when searching the best configuration for Zhuang's variants (i.e., the second row), which is denoted as ``-LN''.}
\label{tab:sup_model_cfg}
\end{table}

For SSL, we follow~\cite{nayebi2021unsupervised} and use specific configurations of optimizer and scheduler for different losses, which are shown in Table~\ref{tab:ssl_loss_cfg}. For SimSiam, we try the CosineAnnealing scheduler both with and without 10 epochs of warmup to search for better model performance. For AE, we use a 3-layer deconvolution network to decode the sparse latent representation from our EAD framework into the original tactile input, regardless of the choice of model architectures.

\begin{table}[h]
\centering
\begin{tabular}{c|ccc}
\toprule
\textbf{Loss} & \textbf{Optimizer} & \textbf{Scheduler} & \textbf{Learning Rate} \\
\midrule
SimCLR & LARS & CosineAnnealing & $10^{\{-1, -2, -3, -4\}}$ \\
SimSiam & SGD & CosineAnnealing (with \& w/o warmup) & $10^{\{-1, -2, -3, -4\}}$ \\
AE & SGD & StepLR & $10^{\{-1, -2, -3, -4\}}$ \\
\bottomrule
\end{tabular}
\vspace{0.05in}
\caption{
\textbf{Optimizer, Scheduler, and Learning Rate Configurations for SSL} when training different model architectures.}
\label{tab:ssl_loss_cfg}
\end{table}

We present the model architectures explored for SSL in Table~\ref{tab:ssl_model_cfg}, where we present the encoder and attender of our EAD architecture, as during the pre-training stage, only the encoder and attender are used, and during the linear probing stage, the decoder is always a one-layer linear classification head.

\begin{table}[h]
\centering


\begin{tabular}{c|c}
\toprule
\textbf{Encoder} & \textbf{Attender} \\
\midrule
Zhuang, Zhuang-\{UGRNN, IntersectionRNN-LN, GRU, LSTM\} & None \\
\midrule
ResNet, Zhuang-IntersectionRNN-LN, Zhuang, S4 & GPT \\
\midrule
ResNet, Zhuang, S4 & Mamba \\
\bottomrule
\end{tabular}

\vspace{0.05in}
\caption{
\textbf{Model Architecture Configurations for SSL}.
We use \{\} to indicate the different variants of encoders. We consider adding layer norm (LN) as a variant when searching the best configuration for Zhuang's variants.
}
\label{tab:ssl_model_cfg}
\end{table}

\newpage

\section{Neural Evaluation}
\label{sec:Neural}

We use the NeuroAI Turing Test \citep{feather2025brain} to evaluate the neural similarity of mice whisking to models performing tactile categorization.

\textbf{RSA Correlation}.
Due to the low number of stimuli, we use RSA \citep{kriegeskorte2008representational} as our correlation metric.
RSA is computed over stimuli and neurons, where the average is over source animals/subsampled source neurons, bootstrapped trials, and train/test splits.
This yields a vector of these average values, which we can take median and s.e.m. over, across animals.

For the neurons of animal $\animalA$ to animal $\animalB$ in the set of animals $\mathcal{A}$ we estimate the RSA correlation:
\begin{equation}\label{rsainteranconid}
\left\langle\rsa\left(\trueAid,\trueBid\right)\right\rangle_{\animalA \in \mathcal{A}: (\animalA,\animalB)\in \mathcal{A}\times\mathcal{A}} \sim \left\langle\dfrac{\rsa\left(\sfAid, \ssBid\right)}{\sqrt{\widetilde{\rsa}\left(\sfAid, \ssAid\right) \times \widetilde{\rsa}\left(\sfBid, \ssBid\right)}}\right\rangle_{\animalA \in \mathcal{A}: (\animalA,\animalB)\in \mathcal{A}\times\mathcal{A}}.
\end{equation}

Each neuron in our analysis is associated with a value for when it was a target animal ($\animalB$), averaged over subsampled source neurons and 1000 bootstrapped trials.
This yields a vector of these average values, which we can take median and standard error of the mean (s.e.m.) over, as we do with standard explained variance metrics.

\textbf{Spearman-Brown Correction}.
The Spearman-Brown correction can be applied to each of the terms in the denominator individually, as they are each correlations of observations from half the trials of the \emph{same} underlying process to itself (unlike the numerator).
\begin{align*}
\widetilde{\rsa}\left(X,Y\right) &\coloneqq \widetilde{\corr}(\rdm(x), \rdm(y)) \\
    &= \frac{2\rsa\left(X,Y\right)}{1 + \rsa\left(X,Y\right)}.
\end{align*}

\textbf{Inter-Animal Consistency}.
To estimate the inter-animal consistency, we evaluate the pooled animal consistency for each animal. One animal is held out at a time, then compared to the pseudo-population aggregated across units from the remaining animals.
We found the mean pooled animal score was 0.175 with a s.e.m. of 0.161 and maximum score of 1.34.

\begin{figure}[h]
  \centering
  
  \subfloat[]{
    \includegraphics[width=0.1\textwidth]{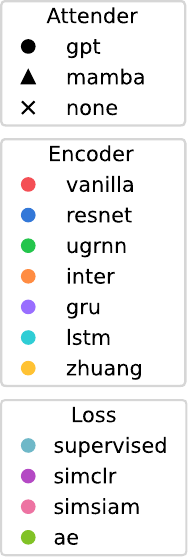} }
  \subfloat[]{
      \includegraphics[width=0.45\textwidth]{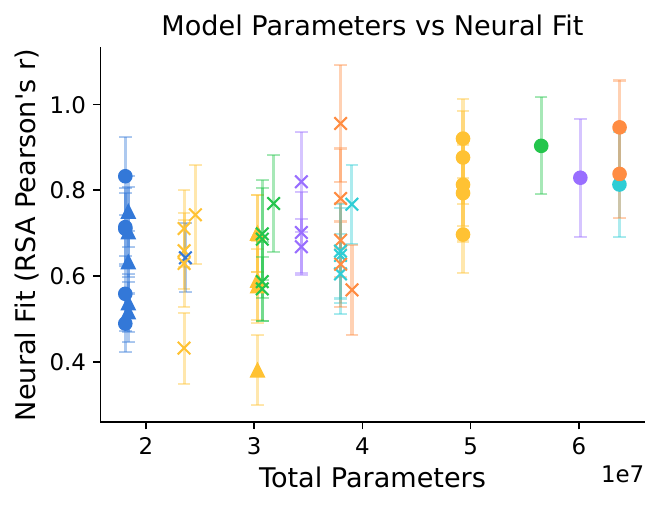}
  }
  \subfloat[]{
      \includegraphics[width=0.45\textwidth]{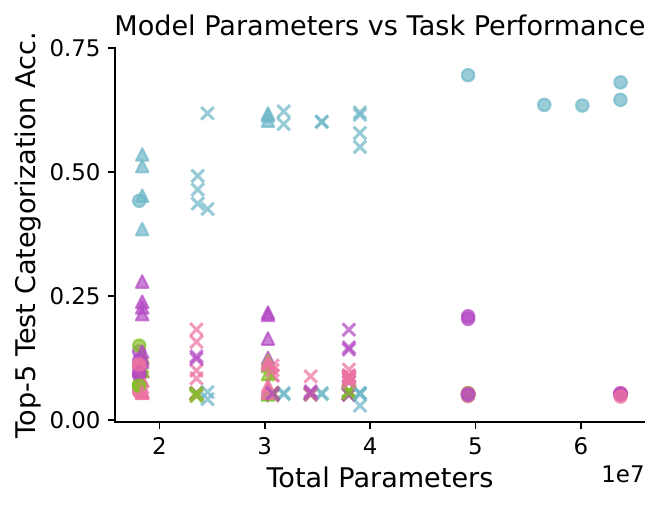}
  }
  \caption{
  \textbf{Model Parameters} compared with categorization performance and neural fit.
  \textbf{(a)} Legends for (b) (Encoder colors) and (c) (Loss colors).
  \textbf{(b)} Models with GPT as an attender have a higher correlation score slightly higher than those without, but high neural fit is still achievable without more parameters as demonstrated by the Inter+SimCLR model (high green ``\green{$\times$}'').
  \textbf{(c)} Models with GPT as the attender has more parameters and, when trained with supervised learning, tends to have higher top-5 categorization accuracy.
  }
  \label{fig:model_params}
\end{figure}
\begin{figure}[h]
  \centering

  \subfloat[]{
      \includegraphics[width=0.63\textwidth]{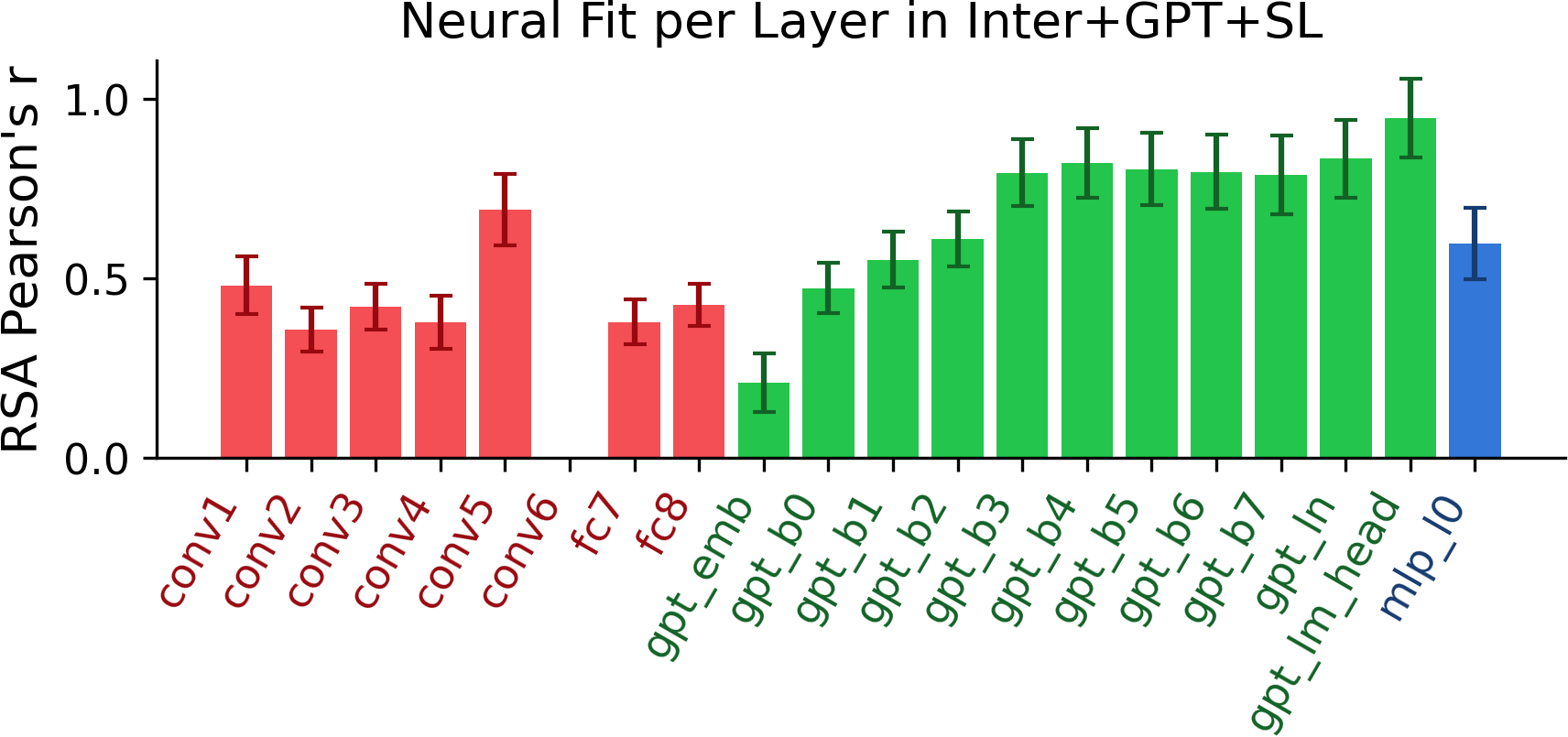}
  }
  \subfloat[]{
      \includegraphics[width=0.37\textwidth]{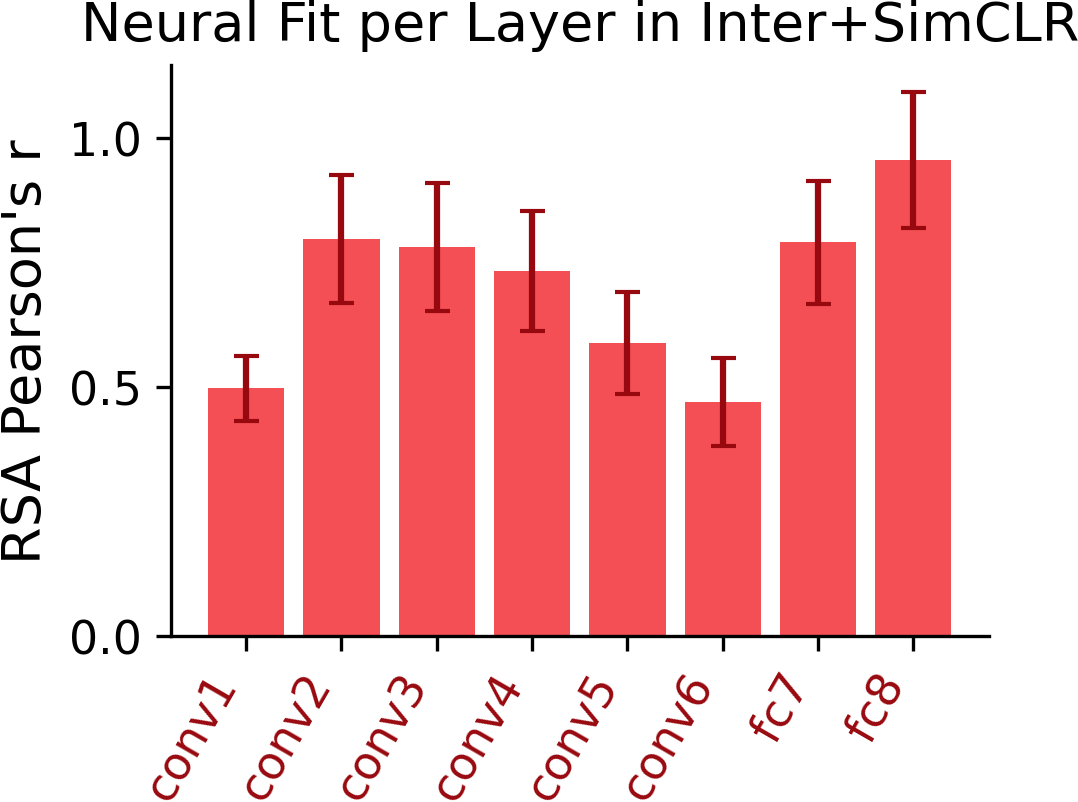}
  }
  \caption{
  \textbf{Neural Fit Score per Model Layers} colored by \red{encoder}, \green{attender}, and \blue{decoder}. No bar means the score is NaN for that layer.
  \textbf{(a)} Inter+GPT+SupervisedLearning is the model that scored the highest neural fit out of the supervised models. We observe that later GPT layers perform increasingly better. 
  \textbf{(b)} The last fully-connected layer of Inter+SimCLR achieved the highest $r$ value.  
  }
  \label{fig:layer_neural_scores}
\end{figure}
\begin{figure}[h]
    \centering
    \subfloat[]{{\includegraphics[width=0.25\textwidth]{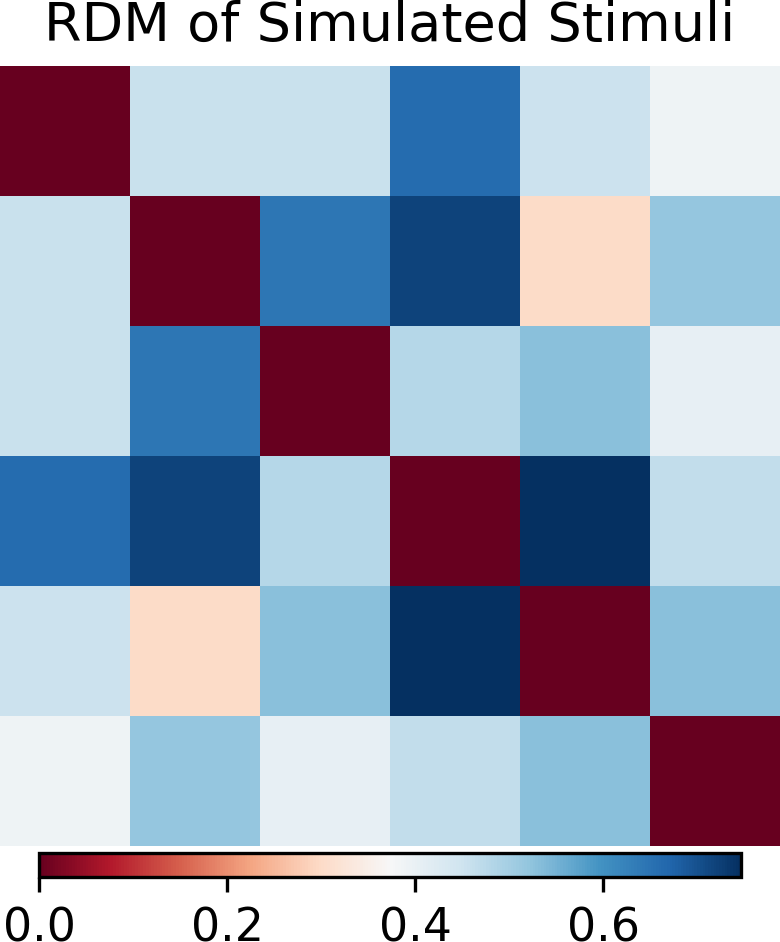} }}
    \subfloat[]{{\includegraphics[width=0.25\textwidth]{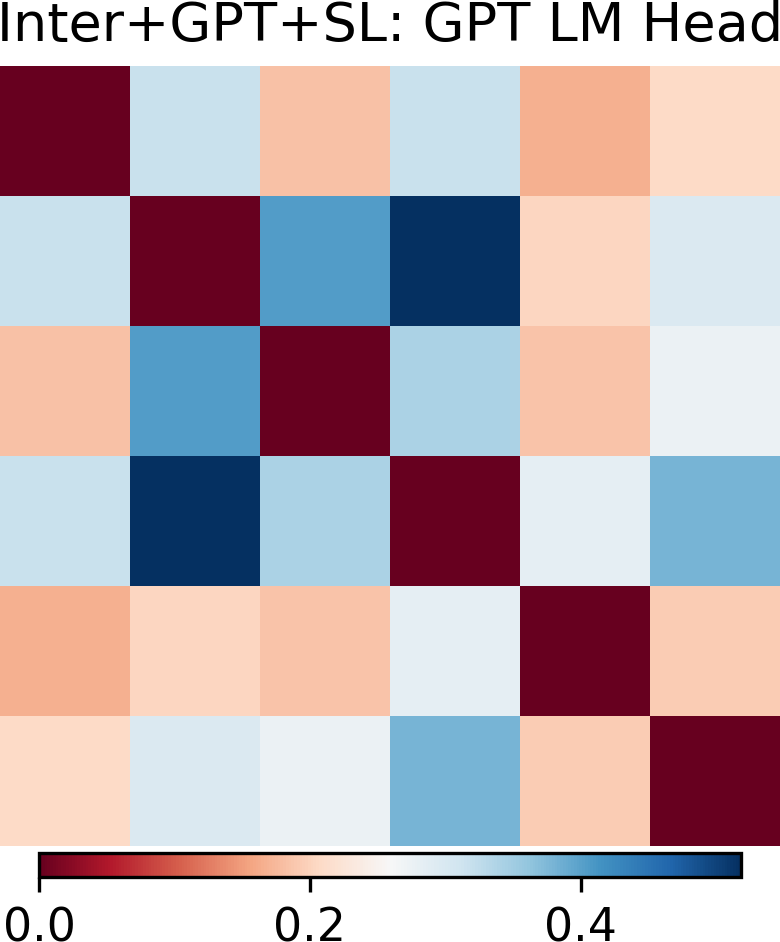} }}
    \subfloat[]{{\includegraphics[width=0.25\textwidth]{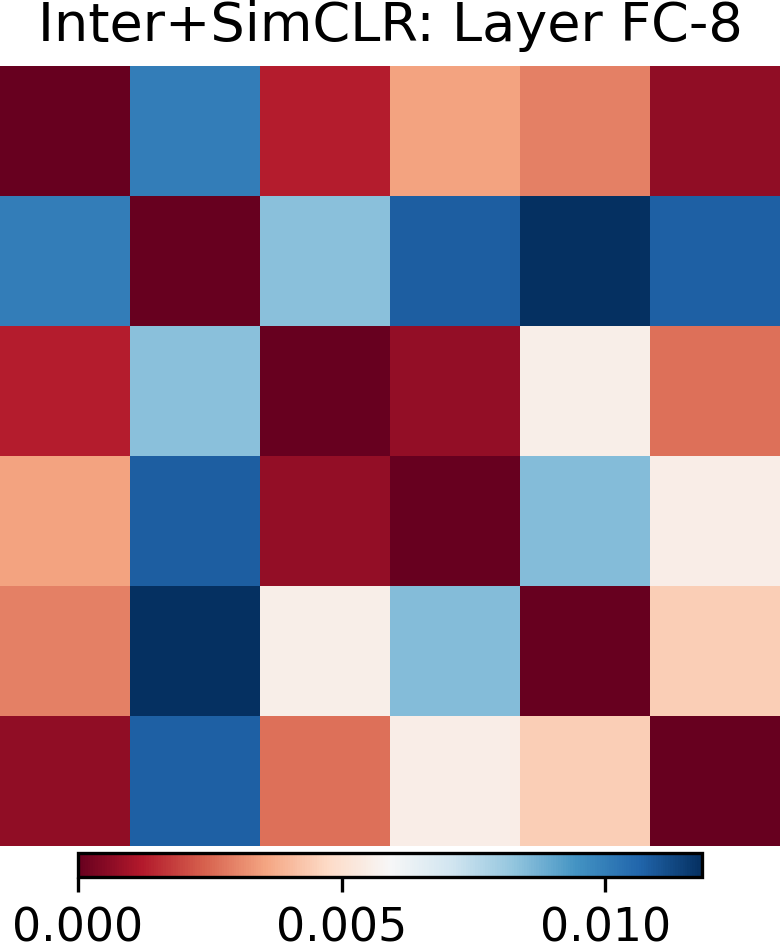} }}
    \subfloat[]{{\includegraphics[width=0.25\textwidth]{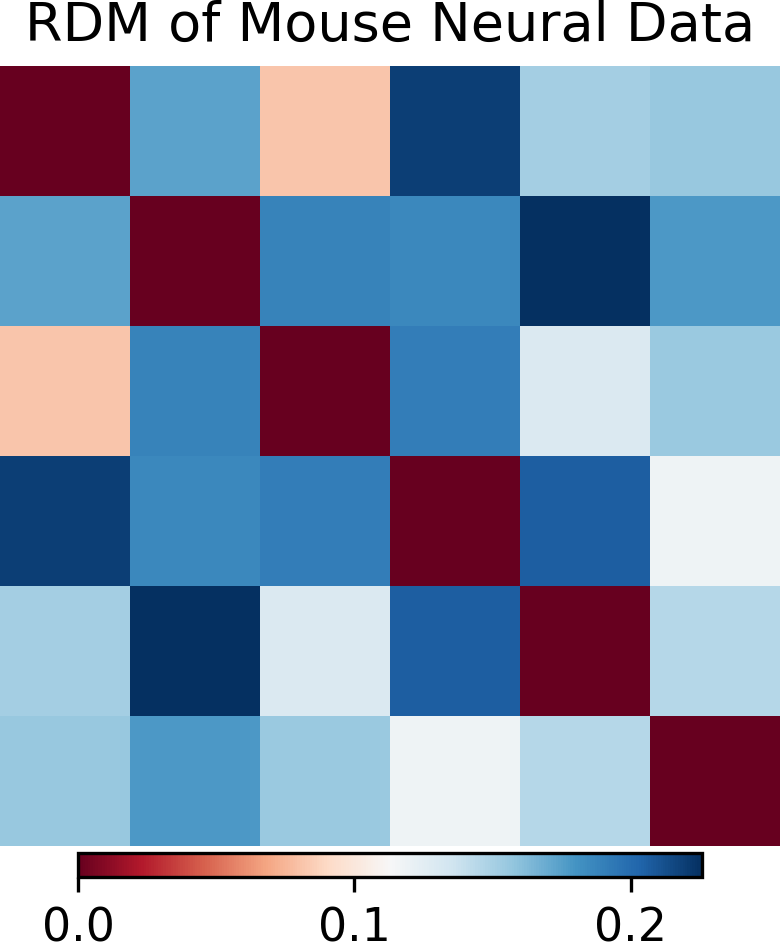} }}
    
    \subfloat[]{
        \begin{minipage}[c]{\textwidth}
            \includegraphics[width=0.67\textwidth]{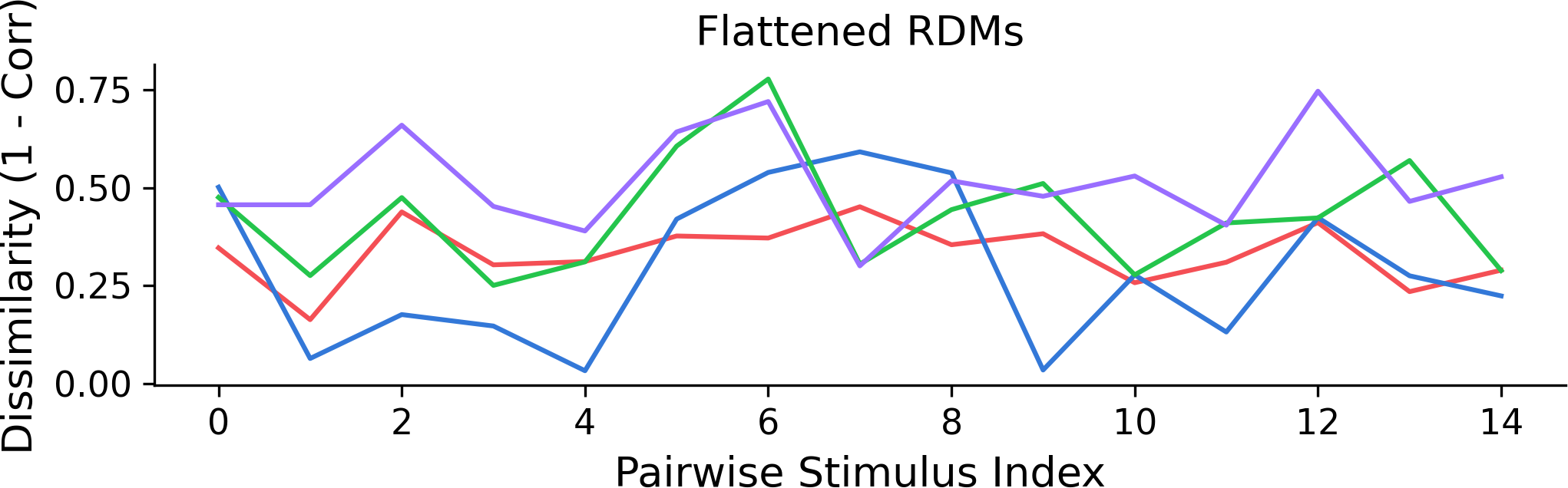}
            \includegraphics[width=0.32\textwidth]{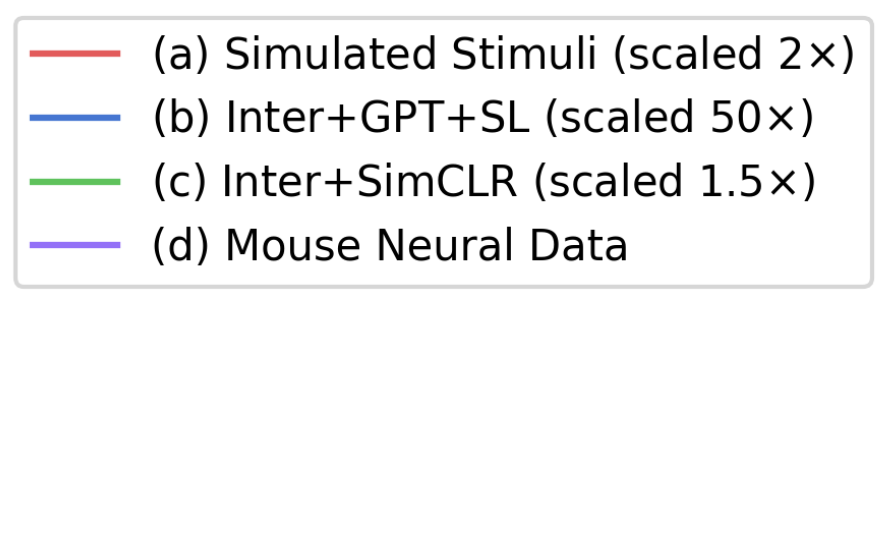}
        \end{minipage}
    }
  \caption{
  \textbf{Representational Dissimilarity Matrices} (RDM) for tactile data and neural evaluation.
  \textbf{(a)} RDM of the 6 simulated stimuli which is used as the model input in neural evaluation.
  \textbf{(b)} The GPT LM Head layer of Inter+GPT+SupervisedLearning is the supervised model with the highest neural fit score.
  \textbf{(c)} The last Fully-Connected layer in Inter+SimCLR achieves the highest neural fit score out of SSL models. 
  \textbf{(d)} RDM performed on the 6 stimuli in the mice neural data.
  \textbf{(e)} A visualization of the flattened RDMs, scaled approximately to fit (RSA correlation does not take scale into account).
  }
\end{figure}

\clearpage

\section{Additional Experiments}
\label{sec:Additional}

As a basic way to probe the distinguishability of the models' learned representations of stimuli, we performed a decoding test with results shown in Table~\ref{tab:stimuli_decoding}.
Note that Inter+SimCLR is the model with the best neural score; Zhuang+GPT+Supervised has the best task score; Inter+GPT+Supervised has best neural score out of supervised models; Resnet+Mamba+SimCLR has best task score out of SSL models.

Rodgers’ task variables (convex/concave) are behavioral probes, and are not necessarily the natural computational goals of somatosensory cortex. Our RSA analysis shows that cortical representational geometry is best matched by tactile-optimized models, especially with self-supervised losses, indicating that the cortex builds a general substrate for tactile recognition and discrimination across diverse objects. When reduced to a six-condition probe, this broad representational space may not project well onto the labels, yielding poor decoding despite underlying ethologically aligned representations.

\begin{table}[h]
\centering
\begin{tabular}{lll}
\hline
\textbf{Model} & \textbf{Shape} (chance=0.5) & \textbf{Distance} (chance=0.33) \\
\hline
Inter+SimCLR             & 0    & 0    \\
Zhuang+GPT+Supervised    & 0.33 & 0.5  \\
Inter+GPT+Supervised     & 0    & 0.5  \\
Resnet+Mamba+SimCLR      & 0    & 0.67 \\
Animal Neural Data       & 0.44 & 0    \\
\hline
\end{tabular}
\vspace{3pt}
\caption{\textbf{Stimuli decoding}. For each of the 6 stimuli, we train a logistic regression model on the 5 other stimuli and test on the 1 held-out stimulus to decode either the shape (convex/concave) or the distance (far/medium/near). 
The stimuli decoding score for Animal Neural Data was obtained by decoding per mouse, then averaging across mice.
}
\label{tab:stimuli_decoding}
\end{table}

We have also tried adding temporal masking, where we randomly mask 75\% of the as a tactile augmentation, but it did not significantly improve the task performance or neural fit score (Table~\ref{tab:masking}).

\begin{table}[h]
\centering
\begin{tabular}{lll}
\hline
\textbf{Model} & \textbf{Top-5 Cat. Accuracy} & \textbf{Neural Fit} \\
\hline
Inter+SimCLR                              & 0.15 & 0.96 \\
Inter+SimCLR \textbf{with Temporal Masking}        & 0.18 & 0.42 \\
Resnet+Mamba+SimCLR                       & 0.23 & 0.70 \\
Resnet+Mamba+SimCLR \textbf{with T. Masking} & 0.28 & 0.67 \\
\hline
\end{tabular}
\vspace{3pt}
\caption{\textbf{Temporal Masking}. Results from retraining two models with temporal masking as an additional tactile augmentation. Top-5 categorization task performance improved slightly (\textasciitilde{}5\%), but neural fit scores decreased.
}
\label{tab:masking}
\end{table}

\end{document}